\documentclass[journal]{IEEEtran}
\usepackage{svg}

\usepackage{cite}

\ifCLASSINFOpdf
\else
\fi
\usepackage{amsmath}
\usepackage{algorithmic}

\usepackage{array}

\ifCLASSOPTIONcompsoc
 \usepackage[caption=false,font=normalsize,labelfont=sf,textfont=sf]{subfig}
\else
 \usepackage[caption=false,font=footnotesize]{subfig}
\fi
\usepackage{dblfloatfix}

\usepackage{bm}

\usepackage{amsfonts}

\begin{document}
%
\title{A Data-driven Probabilistic-based Flexibility Region Estimation Method for Aggregated Distributed Energy Resources}


%
%
%

\author{Mingzhi~Zhang,~\IEEEmembership{Student Member,~IEEE,}
      Xiangqi~Zhu,~\IEEEmembership{Senior Member,~IEEE,}
       and~Ning~Lu,~\IEEEmembership{Fellow,~IEEE.}
\thanks{This project is supported by the DOE ARPA-E EDEGPRO project. 
M. Zhang and N. Lu are with the Department
of Electrical and Computer Engineering, North Carolina State University, Raleigh,
NC, 27606 USA, (e-mail: mzhang33@ncsu.edu, nlu2@ncsu.edu).
Xiangqi Zhu is with the Power Systems Engineering Center, National Renewable Energy Laboratory, Golden, Colorado 80401, USA, (e-mail: xiangqi.zhu@nrel.gov).}
}

%
%

\markboth{Manuscript for review, November~2021}%
{Shell \MakeLowercase{\textit{et al.}}: Bare Demo of IEEEtran.cls for IEEE Journals}

\maketitle

\begin{abstract}

This paper presents a data-driven, distributionally robust chance-constrained optimization method for estimating the real and reactive power controllability of aggregated distributed energy resources (DER). At the DER-level, a two-dimensional flexibility region can be formed based on the real and reactive power regulating limits of each DER considering forecast uncertainty. At the feeder-level, an aggregated flexibility region is computed via a multi-directional search method. In each search direction, extend the real and reactive power of each controllable DER towards its operational limits until: i) all DERs' maximum operational limits are reached or, ii) one or more of the distribution network operational limits are violated. The method enables three key features for operating aggregated DER resources: controllability estimation, visualization, and risk quantification.  Simulation results demonstrated the effectiveness of the algorithm and quantified the impact of different parameter settings on the flexibility region boundary changes. The proposed algorithm is robust and computationally efficient and can meet real-time computing needs. 
\end{abstract}


\IEEEpeerreviewmaketitle

\section{Introduction}
%
%
%
%

\IEEEPARstart{T}{he} renewable energy transition  increases the grid flexibility needs. FERC Order 2222 \cite{FERCOrder2222} enables aggregated distributed energy resources (DER) to compete in all regional organized wholesale electric markets, opening the door for using DERs to provide the much-needed real and reactive power regulations.

Grid operators normally require the following information from a dispatchable flexible resource: an operation point and the operation limits (e.g., upper and lower real and reactive power limits, ramp rates, etc.) in a scheduling window. In this paper, we define the flexibility region as a feasible operation region for an individual DER or an aggregated DER group. Thus, when operated within the flexibility region, two types of constraints must be satisfied: i) the operation limits of each DER are not violated, and ii) the network operational constraints (e.g., nodal voltage or line flow limits) of the distribution system, where the DERs are connected to, will not be violated. 

There are three main challenges when determining the flexibility region for a group of aggregated DERS. First, the number of DERs can be large. To provide MW-level grid services, an aggregator needs to operate hundreds or thousands of DERs with real/reactive power rated from tens of kW/kvar to a few MW/Mvar. Second, the feasibility of a flexibility region is determined not only by the DER-level operational limits but also by the system-level network constraints. Third, DERs have greater operational uncertainty compared with flexible generation resources.

There are four existing approaches for DER operational flexibility estimation: \emph{direct aggregation, computational geometry, Monte Carlo simulation, and optimization}.

The \emph{direct aggregation} based approach \cite{abiri-jahromi_contingency-type_2016,mai_economic_2015,Alharbi2019} normally focuses on a single type of DER (e.g., thermostatically-controlled loads or heating, ventilation, and air conditioning systems) and is not suitable to estimate the operation limits for aggregating a variety of DERs.  

The \emph{computational geometry} based approach \cite{ulbig_analyzing_2015,zhao_geometric_2017,yi2021aggregate} converts the flexibility aggregation problem into a geometric computation problem (e.g. Minkowski sum) so that  the device level operational flexibility regions can be modeled as polygons. However, only the real power limits are computed without considering the potential violation of distribution network operational constraints (e.g., nodal voltage and line flow constraints). Moreover, the combinatorial optimization problem suffers the curse-of-dimensionality issues, especially when aggregating hundreds of heterogeneous DERs. 

The \emph{Monte Carlo simulation} based method\cite{riaz_feasibility_2019} generates a large number of samples within the operation range of each DER. Then, combinations that violate the operational constraints can be identified and excluded. Although the method can effectively account for the DER operational uncertainty, the number of samples required increases exponentially when the number of DER increases, making the computational cost prohibitively higher than the other three approaches.

The \emph{optimization} based approach calculates the aggregated operation limits by optimizing the control of the DERs in order to meet all operational constraints. In \cite{oikonomou_deliverable_2019}, Oikonomou \textit{et. al} proposed a simplified clustered load queuing model to estimate the deliverable real power flexibility limits in the day-ahead market. This method does not consider reactive power operation limits. 
In \cite{silva_estimating_2018}, Silva \textit{et. al}  proposed a searching method to estimate the flexibility limits by identifying a series of bounding nodes for the real and reactive power separately. However, this method does not consider the coupling between the real and reactive power, and the problem formulation is non-convex so the optimality of the solution cannot be guaranteed. An inner-box approximation based optimization method was proposed in \cite{chen_aggregate_2019} by Chen \textit{et. al} to quantify the capacity limits of the real power for aggregated DERs. The authors considered the distribution network model in their problem formulation to address the unbalanced load flows. However, the algorithm considers only the real power limits and requires a distributed algorithm to solve.

The main disadvantage of the aforementioned deterministic, optimization-based approaches is that operational uncertainties inherent in DER operation are not considered. This will inevitable lead to over or under estimation of the DER controllability. 

In \cite{zhao2020active}, Zhao \textit{et. al} proposed a two-stage robust optimization-based method to estimate the maximum feasible real power operation range of virtual power plants.  In \cite{cui2021network}, Cui \textit{et. al} proposed a robust optimization-based formulation to account for the uncertainty in uncontrollable loads. A similar elliptical region-based representation is proposed in \cite{chen2021leveraging } for aggregated real and reactive power flexibility of the distribution system. However, the elliptical-based representation cannot accurately represent the boundary of aggregated two-dimensional flexibility region (as indicated in Fig. 9 of \cite{chen2021leveraging}). A two-stage robust optimization-based method is proposed in \cite{tan2020estimating} for estimating the real and reactive power regulation capability of the virtual power plant considering uncertainties. However, this method does not consider the unbalance power flows in the distribution network.

The main disadvantage of the existing robust optimization-based approaches \cite{zhao2020active, cui2021network, chen2021leveraging, tan2020estimating} is the hedge against the worst-case within a given uncertainty set (e.g. polyhedral/ellipsoidal uncertainty set). This makes the obtained flexibility region overly conservative. In addition, the uncertain set is typically set as a priori with a fixed shape or model that is usually insufficient to capture the complexity of DER operational uncertainties.

In summary, there are three gaps in the existing approaches for identifying the aggregated real and reactive power operational limits in distribution system operation: accounting for the 3-phase unbalanced network operational constraints, considering the close coupling between the real and reactive power, and using the data-driven approach to quantify the DER operational uncertainty. To close those gaps, in this paper, we present a data-driven distributionally robust chance-constrained (DRCC) optimization method for estimating the flexibility region in distribution systems with high-penetration of DERs. The main contributions of this paper are summarized as follows:
\begin{itemize}
\item Provided a rigorous mathematical definition of the flexibility region for typical types of DERs and formulated the physical coupling between the real and reactive power limits considering the DER operational uncertainty.
\item Formulated the 3-phase linearized power flow model into the optimization problem so that the voltage and line flow limits in the unbalanced distribution network are accounted for. 
\item Applied a data-driven method to quantify the uncertainty so that instead of assuming a fixed uncertainty distribution, an ambiguity set of the probability distributions can be constructed from the measurements using statistical inference and data analytics methods.
\item
Developed a DRCC optimization-based algorithm to enable the estimation of the boundary of the aggregated flexibility region boundaries with adjustable risk levels.
\end{itemize}

The rest of this paper is organized as follows. Section \ref{section_device_level_modeling} defines the device-level flexibility region. Section \ref{section_optimization_based_aggregation} introduces the proposed DRCC algorithm for aggregating the device-level flexibility regions. Simulation results are presented in Section \ref{section_case_studies}, and Section \ref{section_conclusion} concludes the paper.

\section{Device-level Flexibility Region Modeling}\label{section_device_level_modeling}
In this section, device-level flexibility region representations are presented and impacts of forecasting errors are discussed.
\subsection{Flexibility Region Representation for Controllable Loads}
\begin{figure}[!h]
	\centering
	\includegraphics[width=2.2 in]{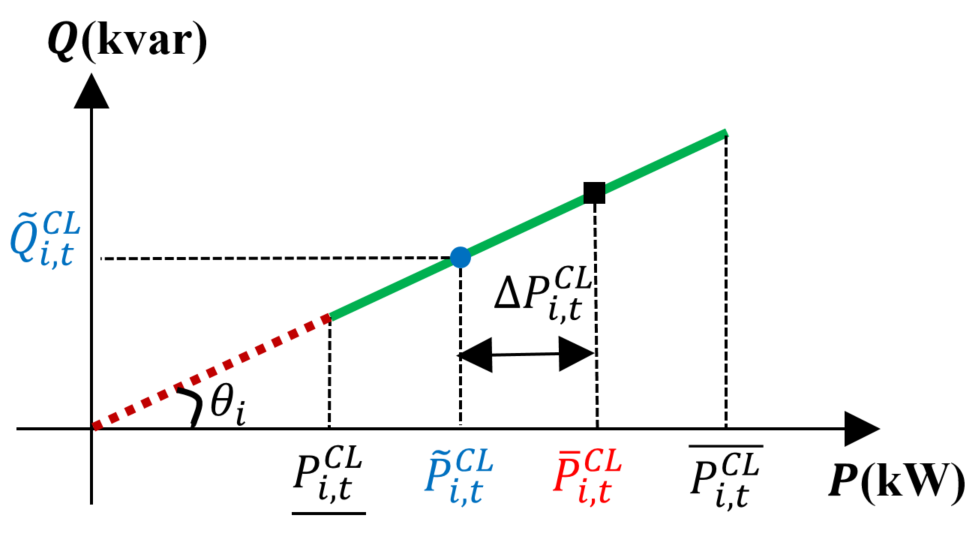}
	\caption{Flexibility region of $i^{\mathrm{th}}$ controllable load.}
    \label{fig_CLFR}
\end{figure}

Assuming there are $N_{\mathrm{CL}}$ controllable loads in a distribution system. At time $t$, controllable load $i$ follows an average power consumption reference, ${\bar{P}_{i,t}}^\mathrm{CL}$. As shown in Fig. \ref{fig_CLFR}, the flexibility of the $i^{\mathrm{th}}$ controllable load is defined as the capability of adjusting its power consumption up or down against ${\bar{P}_{i,t}}^\mathrm{CL}$ in response to a demand response signal, ${\Delta P_{i,t}^\mathrm{C}}$.  Thus, the actual real power consumption of controllable load $i$ at time $t$, $\tilde{P}_{i,t}^{\mathrm{CL}}$, is calculated as:
\begin{equation}\label{equ_controllabe_loads_uncertainty}	
\begin{array}{l}
{\tilde{P}_{i,t}}^{\mathrm{CL}}  = {\bar{P}_{i,t}}^\mathrm{CL} +{\Delta P_{i,t}^\mathrm{CL}},    \forall i \in N_{\mathrm{CL}}.\end{array}
\end{equation}

Assume that the power factor of controllable load $i$, $\cos{\theta_{i}}$, is fixed.  Then, the flexibility region $\mathcal{F}_{i,t}^\mathrm{CL}$ is expressed as:
\begin{equation}\label{equ_feasible_set_loads}
\mathcal{F}_{i,t}^\mathrm{CL}=\left\{\underline{P_{i,t}^\mathrm{CL}} \leq \tilde{P}_{i,t}^\mathrm{CL} \leq \overline{P_{i,t}^\mathrm{CL}},  \tilde{Q}_{i,t}^\mathrm{CL} = \tilde{P}_{i,t}^\mathrm{CL}\tan \theta_{i}   \right\}.
\end{equation}
where $\underline{P_{i,t}^\mathrm{CL}}$ and $\overline{P_{i,t}^\mathrm{CL}}$ is the lower and upper power consumption limits of controllable load $i$ at time $t$, respectively. 

For a controllable load,  $\underline{P_{i,t}^\mathrm{CL}}$ and $\overline{P_{i,t}^\mathrm{CL}}$ are usually known values. Therefore, in this paper, we consider the flexibility region of a controllable load as deterministic.



\subsection{Flexibility Region Representation for PV}
\begin{figure}[!h]
	\centering
	\includegraphics[width=3.0 in]{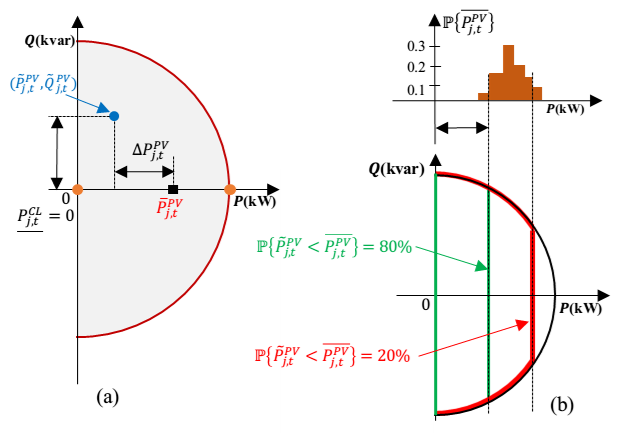}
	\caption{Flexibility region of the $j^{\mathrm{th}}$ PV system.}
    \label{fig_PVFR}
\end{figure}

As shown in Fig. \ref{fig_PVFR}(a), the power output of the $j^{\mathrm{th}}$ PV at time $t$, $\left( \tilde{P}_{j,t}^\mathrm{PV}, \tilde{Q}_{j,t}^\mathrm{PV} \right)$ can be adjusted from its scheduled power output $\left( \bar{P}_{j,t}^\mathrm{PV},\bar{Q}_{j,t}^\mathrm{PV} \right)$, by $\left( {\Delta P_{j,t}^\mathrm{PV}}, {\Delta Q_{j,t}^\mathrm{PV}} \right)$ correspondingly, so we have:
\begin{equation}\label{equ_DER_P_uncertainty}	
\begin{array}{l}
{\tilde{P}_{j,t}}^\mathrm{PV} ={\bar{P}_{j,t}}^\mathrm{PV} +{\Delta P_{j,t}^\mathrm{PV}},   \forall j \in N_{\mathrm{PV}},\\
{\tilde{Q}_{j,t}}^\mathrm{PV} ={\bar{Q}_{j,t}}^\mathrm{PV} +{\Delta Q_{j,t}^\mathrm{PV}},   \forall j \in N_{\mathrm{PV}},
\end{array}
\end{equation}
where $N_\mathrm{PV}$ is the number of PVs in the control group. 

As shown in Fig. \ref{fig_PVFR}(a), to consider both the real and reactive power regulation capability of the $j^{\mathrm{th}}$ PV system, a two-dimensional flexibility region, $\mathcal{F}_{j,t}^\mathrm{PV}$, can be defined as:
\begin{equation}\label{equ_feasible_set_PV}
\mathcal{F}_{j,t}^{\mathrm{PV}} =
\left \{ 
\begin{array}{l}
\underline{P_{j,t}^\mathrm{PV}} \leq \tilde{P}_{j,t}^{\mathrm{PV}} \leq \overline{ P_{j,t}^{\mathrm{PV}} }, \\
\left(\tilde{P}_{j,t}^\mathrm{PV}\right)^2 + \left(\tilde{Q}_{j,t}^\mathrm{PV}\right)^2 \leq \left({S}_{j}^{\mathrm{PV}}\right)^2, 
\end{array}
\right \} 
\end{equation}
where $\underline{P_{j,t}^{\mathrm{PV}}}$ and $\overline{P_{j,t}^{\mathrm{PV}}}$ are minimum and maximum active power output limits of inverter $j$ at time $t$, respectively, and ${S}_{j}^{\mathrm{PV}}$ is the PV rated capacity. If there is a power factor requirement to meet, a power factor constraint $|{\tilde{Q}_{j,t}}^{\mathrm{PV}}| \leq \tan (\theta_j) |{\tilde{P}_{j,t}}^{\mathrm{PV}}|$  can be added to further restrict the feasible operation region. 

In practice, $\underline{P_{j,t}^{\mathrm{PV}}}$ can be set at 0. However, unlike diesel generators, the maximum power output limit of $\overline{P_{j,t}^{\mathrm{PV}}}$ is weather dependent and changes when the solar irradiance varies. As shown in Fig.~\ref{fig_PVFR}(b), at time $t$, the forecasted power output ${\bar{P}_{j,t}}^\mathrm{PV}$ follows a probability distribution, $\mathbb{P}\left\{{\bar{P}_{j,t}}^\mathrm{PV}\right\}$. 
To capture this uncertainty, define $\epsilon_{\mathrm{P}}$ as the risk level, which represents the probability for the PV power output to fall outside the feasible region because of PV forecast errors. Then, we can compute a probabilistic-based flexibility region as:
\begin{equation}\label{equ_PV_cc_constraints}	
\mathbb{P}\left\{ \left(\tilde{P}_{j,t}^{\mathrm{PV}} , \tilde{Q}_{j,t}^{\mathrm{PV}} \right)  \in  \mathcal{F}_{j,t}^{\mathrm{PV}} \right\} \geq 1-\epsilon_{\mathrm{P}} ,  \forall j \in N_{\mathcal{PV}},
\end{equation}
where $\epsilon_{\mathrm{P}}$ is the device-level feasible region violation probability, which is the likelihood of the adjusted operating point is out of the feasible operation region. As shown in Fig. \ref{fig_PVFR} (b), when $\epsilon_{\mathrm{P}}=0.8 $, the flexibility region is bounded by the red line. However, when $\epsilon_{\mathrm{P}}$ is decreased to 0.2, the flexibility region will shrink to the area bounded by the green line, leading to a much more conservative estimation.   

\subsection{Flexibility Region Representation for Battery Energy Storage Systems}
The real and reactive power output of a battery energy storage system (BESS) at time $t$, $\tilde{P}_{k,t}^\mathrm{B}$ and $\tilde{Q}_{k,t}^\mathrm{B}$, can be adjusted from its scheduled outputs, $\bar{P}_{k,t}^\mathrm{B}$ and $\bar{Q}_{k,t}^\mathrm{B}$, by  ${\Delta P_{k,t}^\mathrm{B}}$ and ${\Delta Q_{k,t}^\mathrm{B}}$, respectively, so we have: 
\begin{equation}\label{equ_DER_P_uncertainty}	
\begin{array}{l}
{\tilde{P}_{k,t}}^\mathrm{B} ={\bar{P}_{k,t}}^\mathrm{B} +{\Delta P_{k,t}^\mathrm{B}},   \forall k \in N_{\mathrm{B}},\\
{\tilde{Q}_{k,t}}^\mathrm{B} ={\bar{Q}_{k,t}}^\mathrm{B} +{\Delta Q_{k,t}^\mathrm{B}},   \forall k \in N_{\mathrm{B}},
\end{array}
\end{equation}
where $N_\mathrm{B}$ is the number of BESS in the control group. 

As shown in Fig. \ref{fig_BatteryFR}, the BESS flexibility region, $\mathcal{F}_{j,t}^\mathrm{B}$, is deterministic and can be expressed as:
\begin{equation}\label{equ_feasible_set_battery}
\mathcal{F}_{k,t}^{\mathrm{B}} =
\left \{ 
\left(\tilde{P}_{k,t}^\mathrm{B}\right)^2 + \left(\tilde{Q}_{k,t}^\mathrm{B}\right)^2 \leq \left({S}_{k}^{\mathrm{B}}\right)^2 
\right \}, 
\end{equation}
where ${S}_{k}^{\mathrm{B}}$ is the inverter size of battery storage system.

\begin{figure}[h]\label{fig_BatteryFR}
	\centering
	\includegraphics[width=1.7 in]{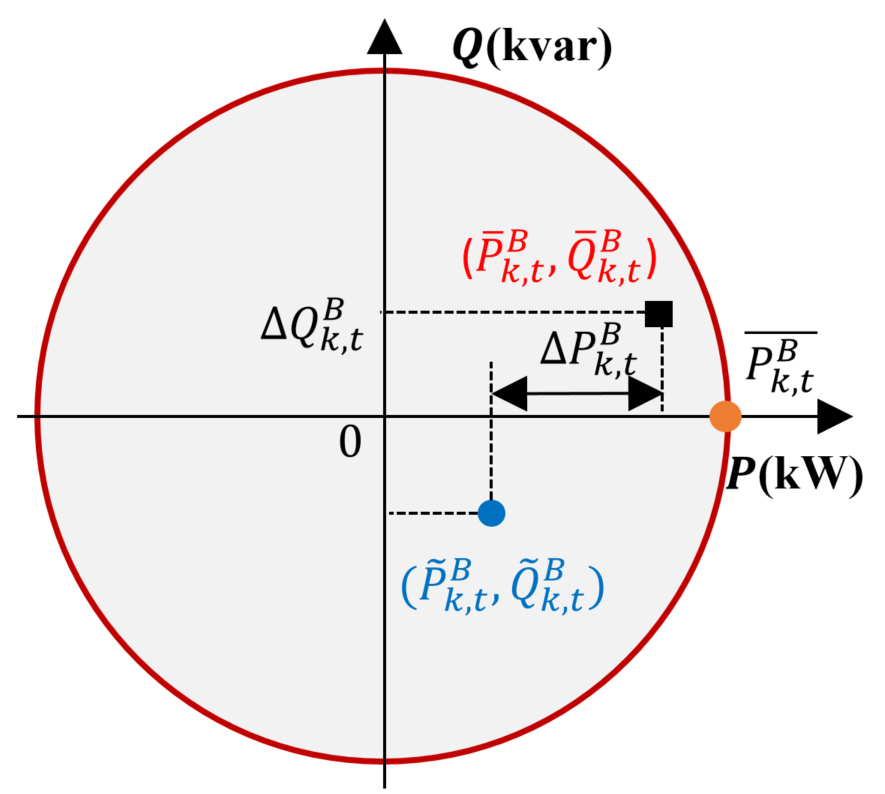}
	\caption{Flexibility region of the $k^{\mathrm{th}}$ BESS.}
    \label{fig_BatteryFR}
\end{figure}

\section{Flexibility Region Aggregation Method}\label{section_optimization_based_aggregation}
In this section, we introduce a distributionally robust optimization method for finding the boundaries of aggregated flexibility region at the feeder head by exploiting the device-level real and reactive power flexibility capacities on different search directions until violations of feeder operation limits (i.e., nodal voltage or line flow limits) are observed. 

\subsection{Feeder-level Flexibility Region Representation}
At the feeder-level, define the actual real and reactive power injections at the time $t$ as $\tilde{P}_{t}^{\mathrm{F}}$ and $\tilde{Q}_{t}^{\mathrm{F}}$, respectively. The actual real and reactive power outputs can be computed by: 
\begin{equation}\label{equ_feeder_PQ}	
\begin{array}{l}
{\tilde{P}_{t}^{\mathrm{F}}} ={\bar{P}_{t}^{\mathrm{F}}} +{\Delta P_{t}^\mathrm{F}},\\
{\tilde{Q}_{t}^{\mathrm{F}}} ={\bar{Q}_{t}^{\mathrm{F}}} +{\Delta Q_{t}^\mathrm{F}},
\end{array}
\end{equation}
where  $\bar{P}_{t}^{\mathrm{F}}$ and $\bar{Q}_{t}^{\mathrm{F}}$ are the feeder scheduled real and reactive power injections and ${\Delta P_{t}^\mathrm{F}}$ and ${\Delta Q_{t}^\mathrm{F}}$ are the corresponding power adjustments. 

To meet the real and reactive power balance requirements at the feeder head, we have:
\begin{equation}\label{equ_feederpower_balance}	
\begin{aligned}
	\bar{P}_{t}^F=& \sum\limits_{i=1}^{N_{\mathrm{CL}}}{\bar{P}_{i,t}}^{\mathrm{CL}}  + \sum\limits_{j=1}^{N_{\mathrm{PV}}}{\bar{P}_{j,t}}^\mathrm{PV} + \sum\limits_{k=1}^{N_{\mathrm{B}}}{\bar{P}_{k,t}}^\mathrm{B}    + \sum\limits_{m=1}^{N_{\mathrm{NCL}}}{P_{m,t}^{\mathrm{NCL}}} + P_{t}^{\mathrm{loss}},\\
	\bar{Q}_{t}^F=& \sum\limits_{i=1}^{N_{\mathrm{CL}}}{\bar{Q}_{i,t}}^{\mathrm{CL}}  + \sum\limits_{j=1}^{N_{\mathrm{PV}}}{\bar{Q}_{j,t}}^\mathrm{PV}  + \sum\limits_{k=1}^{N_{\mathrm{B}}}{\bar{Q}_{k,t}}^\mathrm{B}  + \sum\limits_{m=1}^{N_{\mathrm{NCL}}}{Q_{m,t}^{\mathrm{NCL}}} + Q_{t}^{\mathrm{loss}},\\
\end{aligned}	
\end{equation}
where $P_{m,t}^{\mathrm{NCL}}$ and $Q_{m,t}^{\mathrm{NCL}}$ are the forecasted real and reactive power consumption of the non-controllable load $l$, and $P_{t}^{\mathrm{loss}}$ and $Q_{t}^{\mathrm{loss}}$ are feeder real and reactive power losses. 

The available feeder-level flexibility capacity ${\Delta P_{t}^\mathrm{F}}$ and ${\Delta Q_{t}^\mathrm{F}}$ is achieved through the aggregation of device-level flexibility capacity by regulating the active and reactive power of controllable loads and inverters without violating any feeder operational constraints:
\begin{equation}\label{equ_DeltaPQ_balance}	
\begin{aligned}
	{\Delta P_{t}^\mathrm{F}}=& \sum\limits_{i=1}^{N_{\mathrm{CL}}}{{\Delta P_{i,t}^\mathrm{CL}}}  + \sum\limits_{j=1}^{N_{\mathrm{PV}}}{\Delta P_{j,t}^\mathrm{PV}}+ \sum\limits_{k=1}^{N_{\mathrm{B}}}{\Delta P_{k,t}^\mathrm{B}},\\
{\Delta Q_{t}^\mathrm{F}}=& \sum\limits_{i=1}^{N_{\mathrm{CL}}}{{(\Delta P_{i,t}^\mathrm{CL}}\tan \theta_{i})}  + \sum\limits_{j=1}^{N_{\mathrm{PV}}}{\Delta Q_{j,t}^\mathrm{PV}}+ \sum\limits_{k=1}^{N_{\mathrm{B}}}{\Delta Q_{k,t}^\mathrm{B}},
\end{aligned}	
\end{equation}

Therefore, the two-dimensional feeder-level flexibility region $\mathcal{F}_{t}^{\mathrm{F}}$ at the time $t$ can be defined as:
\begin{equation}\label{equ_feeder_feasible_set}
\mathcal{F}_{t}^{\mathrm{F}} =
\left \{ 
\begin{array}{l}
\underline{P_{t}^{\mathrm{F}}} \leq {\tilde{P}_{t}}^{\mathrm{F}} \leq \overline{P_{t}^{\mathrm{F}}}, \\
\underline{Q_{t}^{\mathrm{F}}} \leq {\tilde{Q}_{t}}^{\mathrm{F}} \leq \overline{Q_{t}^{\mathrm{F}}}, 
\end{array}
\right \} 
\end{equation}
where ($\underline{P_{t}^\mathrm{F}}$, $\overline{P_{t}^\mathrm{F}}$) and ($\underline{Q_{t}^\mathrm{F}}$, $\overline{Q_{t}^\mathrm{F}}$) stand for the lower and upper limits of feeder real and reactive power at time $t$ respectively.

\subsection{Nodal Voltage and Line Flow Estimation Method}\label{linearized_PF}
The boundary of the feeder-level flexibility region can be determined by perturbing the operating point of flexibility assets until violations of feeder network operation constraints (e.g., nodal voltage and line flow limits) are observed.  A distribution feeder can have tens of (or even hundreds of) DERs and controllable loads participating in the flexibility capacity provisions. Thus, a detailed, 3-phase unbalanced power network model is required for calculating nodal voltage and line flow changes caused by the device-level flexibility capacity realizations.

The non-linearity of exact full AC power flow formulation making it computationally expensive when applying the proposed optimization-based flexibility region estimation method. Therefore, we applied fixed-point linearized power flow  (a sensitivity matrix based approach) introduced in \cite{Bernstein2018} by Bernstein \textit{et. al} to shorten the computing time and the complexity of the optimization problem formulation. For a multi-phase, unbalanced distribution network with $N$ buses and $L$ lines, the nodal voltage vector, $\tilde{\mathbf{V}}$, is calculated as:
\begin{equation}\label{equ_linear_model_voltage}
\tilde{\mathbf{V}}= \mathbf{M}_{Y} \mathbf{x}_{Y}+\mathbf{M}_{\Delta} \mathbf{x}_{\Delta}+\mathbf{\alpha}, 
\end{equation}
where $\mathbf{x}_{Y}=\left[\mathbf{p}_{Y}^{\top},\mathbf{q}_{Y}^{\top}\right]^{\top}$ and 
$\mathbf{x}_{\Delta} = \left[\mathbf{p}_{\Delta}^{\top},\mathbf{q}_{\Delta}^{\top}\right]^{\top}$, each with a dimension of $\mathbb{D}^{6N \times 1}$, are the 1-phase nodal real and reactive power injection vectors of the $Y/\Delta$ connected sources, 
$\mathbf{M}_{Y/\Delta}$ with a dimension of $\mathbb{D}^{3N \times 6N}$ is the voltage sensitivities matrix of $Y/\Delta$ connected sources, and $\mathbf{\alpha}$ with a dimension of $\mathbb{D}^{3N \times 1}$ is the no-load nodal voltage vector.



Using the method introduced in \cite{christakou2013efficient}, the branch current $\tilde{\mathbf{I}}$ with respect to the nodal
power injections in the network, can be calculated as:
\begin{equation}\label{equ_linear_model_current}
\tilde{\mathbf{I}}=  \mathbf{N}_{Y} \mathbf{x}_{Y}+\mathbf{N}_{\Delta} \mathbf{x}_{\Delta}+\mathbf{\beta},
\end{equation}
where $\mathbf{N}_{Y/\Delta}$ is current sensitivities matrix of the $Y/\Delta$ connected sources with a dimension of $\mathbb{D}^{3L \times 6N}$.

This linearization method differs from the local approximation approach (e.g., the first-order Taylor method) by providing an interpolation between specific loading conditions and the no load condition to achieve better global approximation accuracy. Please refer to \cite{Bernstein2018} and \cite{christakou2013efficient} for more details.

\subsection{Problem Formulation}\label{DRCC_formulation}
To estimate the feeder-level aggregated flexibility region boundary, a search based optimization problem is formulated to estimate the flexibility region for regulating the DERs without violating any feeder operational constraints. The search directions are defined as: 
\begin{equation}\label{equ_search_direction}
\lambda_{P,k}^2 + \lambda_{Q,k}^2 = 1,
\end{equation}
where $\lambda_{P,k}$ and $\lambda_{Q,k}$ are the active and reactive power searching parameters that define a search direction $k$. 

\begin{figure}[!h]
	\centering
	\includegraphics[width=2.6in]{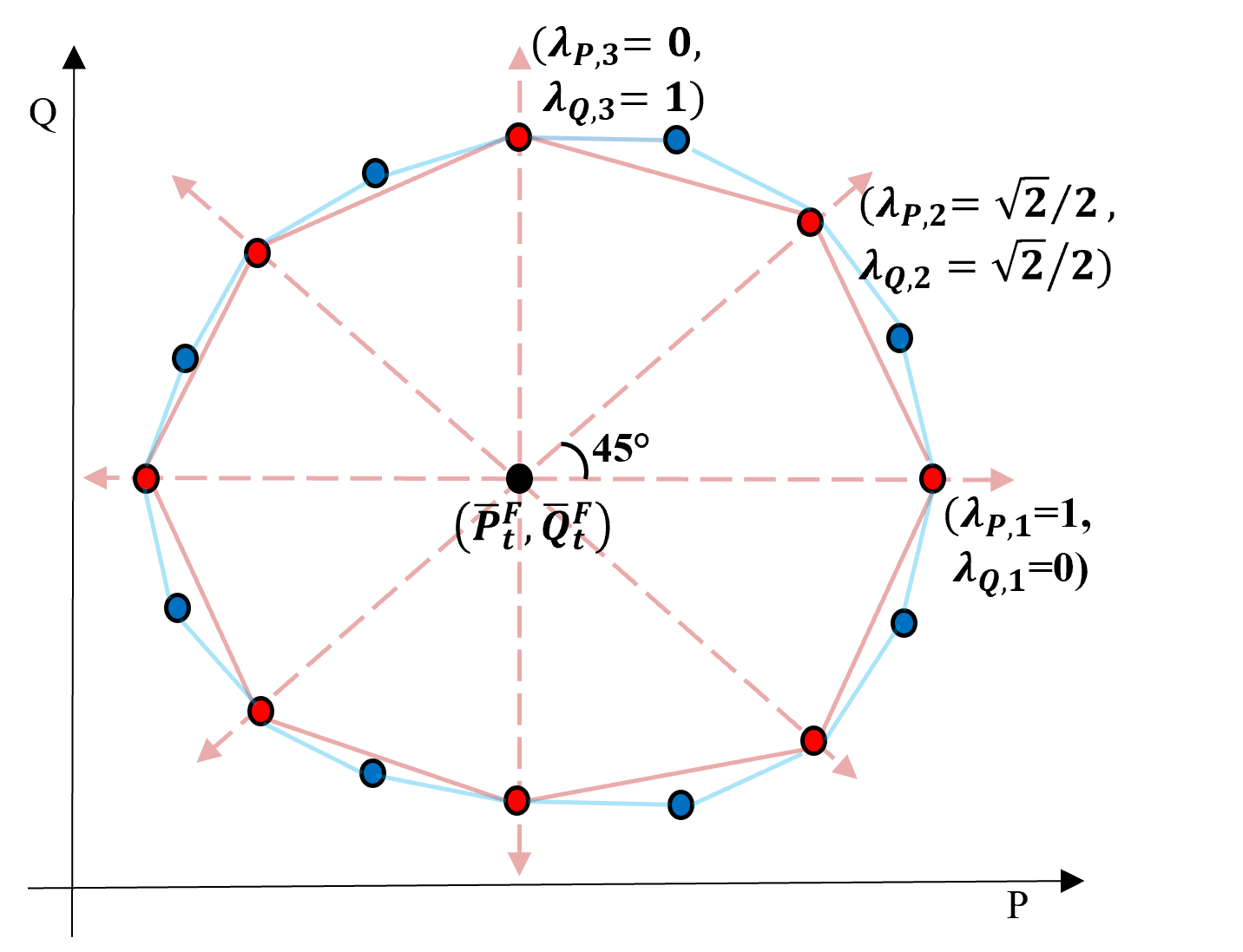}
	\caption{Polygon-based flexibility region bounded by power injection limits calculated from search directions. (Red dots representing the 8-direction case; red and blue dots representing the 16-direction case). }
    \label{fig_search_direction}
\end{figure}

As shown in Fig. \ref{fig_search_direction}, if the total search direction number, $k$, is set as 8,  we can find 8 pairs of $\tilde{P}_{t,k}^{F*}$ and $\tilde{Q}_{t,k}^{F*}$ that represents the feeder active and reactive power injection limits, each corresponding to a different search direction, within which no voltage and current limits will be violated. By connecting the 8 bounding points (red dots) with straight lines, the aggregated feeder-level flexibility region can be approximated by an octagon. If we increase the search direction number to 16, an extra set of 8 dots (blue dots) will be identified on the flexibility region boundary. Thus, the flexibility region boundary can be more accurately approximated using a polygon when the search direction $k$ increases. 

The objective function of the boundary operating points identification problem for a specific search direction $k$ can be formulated as:
\begin{equation}\label{equ_max_obj}
\max \quad \left(\lambda_{P,k}* \tilde{P}_{t,k}^{F} + \lambda_{Q,k} * \tilde{Q}_{t,k}^{F} \right),
\end{equation}
where ($\tilde{P}_{t,k}^{F}$, $\tilde{Q}_{t,k}^{F}$) is the feeder-level operating point at time $t$ for search direction $k$. For example, if $k=1$, we have $\lambda_{P,1}=1$ and $\lambda_{Q,1}=0$, based on (\ref{equ_search_direction}), objective function (\ref{equ_max_obj}) becomes $\max \left( \tilde{P}_{t,1}^{F} \right) $, which is to maximize the feeder real power consumption without violating any operational constraints. 

The objective function has the following constraints: 
\subsubsection{Power balancing constraints}
The real and reactive power balancing constraints are defined in (\ref{equ_feeder_PQ}), (\ref{equ_feederpower_balance}) and (\ref{equ_DeltaPQ_balance}). Those constraints need to be satisfied for all search directions to ensure that the flexibility results obtained are in accordance with the power flow results.

\subsubsection{Device-level operational chance constraints} 
As introduced in section \ref{section_device_level_modeling}, for each flexibility asset, when providing the flexibility services, the available capacity is constrained by its device-level feasible region, defined in (\ref{equ_feasible_set_loads}), (\ref{equ_PV_cc_constraints}) and (\ref{equ_feasible_set_battery}).


\subsubsection{System-level operational chance constraints}
The feeder-level real and reactive power regulation flexibility can be achieved by aggregating the device-level flexibility of each individual DER (i.e., controllable loads, PVs, and battery systems) without violating the voltage or line flow constraints. To make sure the system operational constraints are met with high probability in the flexibility aggregation process, the system-level operational chance constraints on nodal voltage and line flow are defined as: 
\begin{subequations}\label{equ_system_cc_constraints}	
\begin{align}
\mathbb{P}\left\{\tilde{V}_{n,t} \leq V_{n}^{\max }\right\} \geq 1-\epsilon_{V}, & \forall_{n \in \mathcal{N}}  \label{equ_system_cc_constraints_Vmax}	\\
\mathbb{P}\left\{\tilde{V}_{n,t} \geq V_{n}^{\min }\right\} \geq 1-\epsilon_{V}, & \forall_{n \in \mathcal{N}}   \label{equ_system_cc_constraints_Vmin}	\\
\mathbb{P}\left\{\tilde{I}_{l,t} \leq I_{l}^{\max }\right\} \geq 1-\epsilon_{I}, & \forall_{l\in \mathcal{L}}   \label{equ_system_cc_constraints_Imax}
\end{align}	
\end{subequations}
where $\tilde{V}_{n,t}$ and $\tilde{I}_{l,t}$ are nodal voltage and line flow current of node $n$ and line $l$ corresponding to the current system operating point. $\tilde{V}_{n,t}$ and $\tilde{I}_{l,t}$ can be calculated using the linearized power flow equations defined by (\ref{equ_linear_model_voltage}) and (\ref{equ_linear_model_current}),  $\epsilon_{V}$ and $\epsilon_{I}$ are likelihoods of voltage and line flow constraints violation.

\subsection{Data-driven Distributional Robust Reformulation}\label{DRCC_section}
The main challenge in solving chance constrained programming problem is that the probability distribution function (PDF) of uncertain variables are rarely available in practice. Relaying on a predetermined PDF, such as Gaussian \cite{lubin2015robust,mieth2018data} or WeilBull \cite{liu2010economic}, can lead to sub-optimal results \cite{smith2006optimizer}. The PDF can only be estimated from a finite number of uncertainty realizations using the data-driven approach. 

Instead of using a single type of PDF, the proposed data-driven DRCC optimization formulates a family of PDFs (i.e., an ambiguity set) that satisfy the statistical characteristics of each stochastic variable using the following generalized form: 
\begin{equation}\label{DRCC}	
\inf _{\mathbb{P}_{\xi} \in \mathcal{D}_{\xi}} \mathbb{P}_{\xi}(f(x, \xi) \leq 0) \geq 1-\epsilon,
\end{equation}
where $x$ represents the decision variables, $\xi$ is a random variable following PDF $\mathbb{P}_{\xi}$ that belongs to an ambiguity set $\mathcal{D}_{\xi}$, and $\epsilon$ is the pre-defined risk level. DRCC enforces the satisfaction of the chance constraints for all the PDFs in the ambiguity set.

The choice of the ambiguity set is critical for the formulation of the data-driven DRCC. One commonly used approach is the moment-based ambiguity set introduced in \cite{zhang2016distributionally,chen2016data,shi2018distributionally} where the first and second-order based moment information are extracted from the historical uncertain variables. The ambiguity set is expressed as:
\begin{equation}\label{ambiguity set}	
\mathcal{D}_{\xi}=\left\{\mathbb{P}_{\xi} \in \mathcal{P}^{\prime}: \mathbb{E}_{\mathbb{P}_{\xi}}[\xi]=\mu, \mathbb{E}_{\mathbb{P}_{\xi}}\left[\xi \xi^{\top}\right]=\Sigma\right\},
\end{equation}
where $\mu$ and $\Sigma$ are the mean and covariance of the uncertain variables that can be obtained using statistical inference of historical uncertainty realizations.

Based on \cite{calafiore2006distributionally}, the chance constraint (\ref{DRCC}) can be reformulated into a second-order cone constraint:
\begin{equation}\label{DRCC_reformulation}
\mu_{f(x, \xi)}  + K_{\epsilon} * \sigma_{f(x, \xi)} \le 0,
\end{equation}
where $\mu_{f(x, \xi)}$ is the mean value of the constraint, $\sigma_{f(x, \xi)}$ is the standard deviation of the constraint, and $K_{\epsilon} = \sqrt{\left(1-\epsilon \right)/\epsilon} $ is a adjustable coefficient, controlling the robustness of the chance
constraint.

To account for the operational uncertainty in PV outputs, the device-level chance constraints of the PV defined in (\ref{equ_PV_cc_constraints}) can be reformulated as:
\begin{equation}\label{equ_DRCC_PV}
\begin{array}{c}
0 \leq \tilde{P}_{j,t}^{\mathrm{PV}} \leq  \mu_{\bar{P}_{j,t}^\mathrm{PV}} - K_{\epsilon_P} * \sigma_{\bar{P}_{j,t}^\mathrm{PV}}, \\
\left(\tilde{P}_{j,t}^\mathrm{PV}\right)^2 + \left(\tilde{Q}_{j,t}^\mathrm{PV}\right)^2 \leq \left({S}_{j}^{\mathrm{PV}}\right)^2, \forall j \in N_{\mathcal{PV}},
\end{array}
\end{equation}
where $\mu_{\bar{P}_{j,t}^\mathrm{PV}}$ is the expectation of power output of PV $j$ at $t$, $\sigma_{\bar{P}_{j,t}^\mathrm{PV}}$ is the standard deviation of the forecasted power output, and $K_{\epsilon_P}= \sqrt{\left(1-\epsilon_p \right)/\epsilon_P}$ is the corresponding coefficient to control the robustness against forecast errors.

To account for uncertainty propagation, i.e., the uncertainty in the nodal injection estimation propagates to the calculation of the nodal voltages and line flows, we reformulate the system-level operational chance constraints as (\ref{equ_system_cc_constraints}) can be reformulated as:
\begin{subequations}\label{equ_system_DRCC}	
\begin{align}
\mu_{\tilde{V}_{n,t}} + K_{\epsilon_V} * \sigma_{\tilde{V}_{n,t}} \le V_{n}^{\max}, & \forall_{n \in \mathcal{N}}\\
\mu_{\tilde{V}_{n,t}} - K_{\epsilon_V} * \sigma_{\tilde{V}_{n,t}} \ge V_{n}^{\min}, & \forall_{n \in \mathcal{N}}\\
\mu_{\tilde{I}_{l,t}} + K_{\epsilon_I} * \sigma_{\tilde{I}_{l,t}} \le I_{l}^{\max}, & \forall_{l\in \mathcal{L}} 
\end{align}	
\end{subequations}
where $\mu_{\tilde{V}_{n,t}}$ and $\sigma_{\tilde{V}_{n,t}}$ are the mean and standard deviation values of node $n$ voltage, $\mu_{\tilde{I}_{l,t}}$ and $\sigma_{\tilde{I}_{l,t}}$ are the mean and standard deviation values of line $l$ current, and $K_{\epsilon_V},K_{\epsilon_I}$ are the corresponding voltage and line constraint coefficients.

After that, the search based chance constrained optimization can be reformulated into a DRCC optimization problem by:
\begin{subequations}\label{equ_DRCC}	
\begin{align}
&\text{Objective function:~}  (\ref{equ_max_obj})\\
&\text{Power balance constraints:~} (\ref{equ_feeder_PQ}), (\ref{equ_feederpower_balance}) \, \text{and} \, (\ref{equ_DeltaPQ_balance})\\
&\text{Device-level operational constraints:~} (\ref{equ_feasible_set_loads}),(\ref{equ_feasible_set_battery}) \, \text{and} \, (\ref{equ_DRCC_PV})\\
&\text{System-level operational constraint:~} (\ref{equ_system_DRCC})
\end{align}	
\end{subequations}

\section{Numerical Studies}\label{section_case_studies}
\begin{figure}[!h]
	\centering
	\includegraphics[width=3.2in]{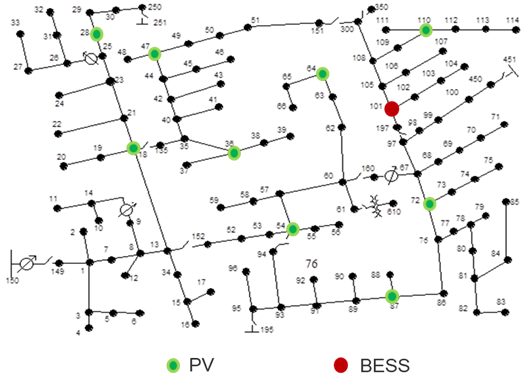}
	\caption{Modified IEEE-123 bus system. }
    \label{fig_IEEE123}
\end{figure}

As shown in Fig. \ref{fig_IEEE123}, 9 PV systems and 1 BESS are added to the IEEE 123-bus system to demonstrate the proposed algorithm. The parameters and connection points of the DERs are listed in Table \ref{tab:flexibility_assets}. Each load node in the 123-bus system is assigned a unique 5-minute load profile using the method introduced in \cite{wang2020data}. Note that the load profiles are derived from the 1-minute Pecan Street Dataset \cite{pecan_street}. The PV power outputs are derived from the NREL solar irradiance data \cite{NREL_solar}. Because the supply radius of a distribution feeder is within twenty miles, we assume that all PVs follow the same power output shape.

\begin{table}[h]
\caption{DER location and Power rating}
\label{tab:flexibility_assets}
\centering
\begin{tabular}{cccc}
\hline \hline
\begin{tabular}[c]{@{}c@{}}DER Assets\end{tabular} &
  \begin{tabular}[c]{@{}c@{}}Location (Phase)\end{tabular} &
  \begin{tabular}[c]{@{}c@{}} Capacity\end{tabular} &
  \begin{tabular}[c]{@{}c@{}}Inverter Size\end{tabular} \\ \hline 
PV 1     & 18   (a,b,c) & 250   KW  & 250   KVA \\
PV   2   & 28   (a,b,c) & 60   KW   & 60   KVA  \\
PV   3   & 36 (b)        & 40   KW   & 40   KVA  \\
PV   4   & 47 (a,b,c)    & 100   KW  & 100   KVA \\
PV   5   & 54 (a,b,c)    & 300   KW  & 300   KVA \\
PV   6   & 64 (a,b,c)    & 80   KW   & 80   KVA  \\
PV   7   & 72 (a)        & 60   KW   & 60   KVA  \\
PV   8   & 87 (a,b,c)    & 120   KW  & 120   KVA \\
PV   9   & 110 (a)       & 80   KW   & 80   KVA  \\
BESS 1   & 101 (a,b,c)   & 100  KWh  & 100   KVA \\ \hline \hline
\end{tabular}
\end{table}

The feeder-head real and reactive net load profiles for three typical day types (sunny, cloudy, and overcast) are shown in Fig. \ref{fig_feeder_power}. To produce these net load profiles, we keep the nodal load profiles the same while varying the PV profiles for different day types. Because we assume that at the base operation point, the PVs and the BESS are operated under the unity power factor mode, the feeder reactive power curves are nearly identical for different day type. In a sunny day, the net load valley occurs around the noontime where the PV power output is the highest and the net load peak occurs between 19:00-20:00 around the sunset. In a cloudy day, the PV outputs fluctuates quickly during the daytime. In an overcast day, the net load curve is significantly higher than that of the sunny day due to the low PV power outputs. In subsequent sections, the impact of different day types on the flexibility region estimation will be addressed.
\begin{figure}[!h]
	\centering
	\includegraphics[width=2.8 in]{feeder_power_variaton_weathers.pdf}
	\caption{Feeder-head load profiles in sunny, overcast and cloudy days.}
    \label{fig_feeder_power}
\end{figure}



\subsection{Time-series-based Aggregated Flexibility Region Analysis without Uncertainty Considerations}
In this case, the proposed flexibility region aggregation algorithm is executed every 30-minute using measured nodal load and PV outputs. Thus, each flexibility region is a snapshot of the operation boundary at the beginning of the 30-minute interval. Thus, for a 24-hour operation window, 48 feeder-head flexibility regions are calculated.

\begin{figure}[!h]
	\centering
	\includegraphics[width=3.3 in]{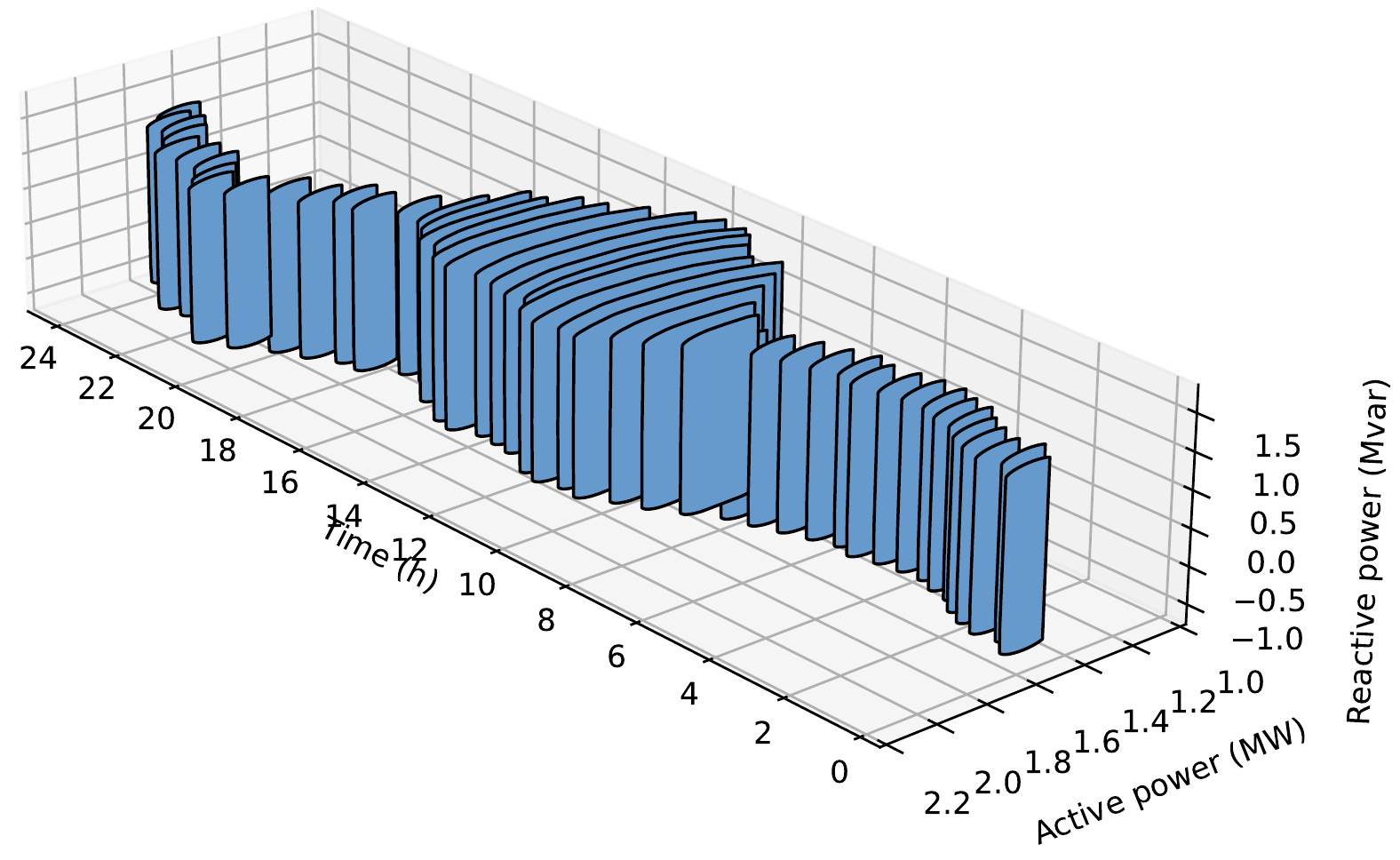}
	\caption{Time-series plots of the feeder-level flexibility region for a 24-hour period (DERs considered include the PV and battery systems).}
    \label{fig_flex_timeseries}
\end{figure}

In the first scenario, we considered only inverter-based resources, i.e., PVs and batteries, as listed in Table \ref{tab:flexibility_assets}. As shown in Fig. \ref{fig_flex_timeseries}, at night hours (i.e., 0:00-8:00 and 16:00-24:00), the PV active power outputs are zero so the feeder real power operation adjustment is achieved by battery systems. Thus, the flexibility is limited along the active power direction. During the sunlight hours (i.e., from 8:00 to 16:00), the active power flexibility will increase when PVs start to generate. Because PVs can provide reactive power continuously throughout the day (see Fig. \ref{fig_PVFR}), the upper and lower limits of the reactive power regulation regions do not vary dramatically during a day. 





\begin{figure}[h]
	\centering
	\includegraphics[width=3.3 in]{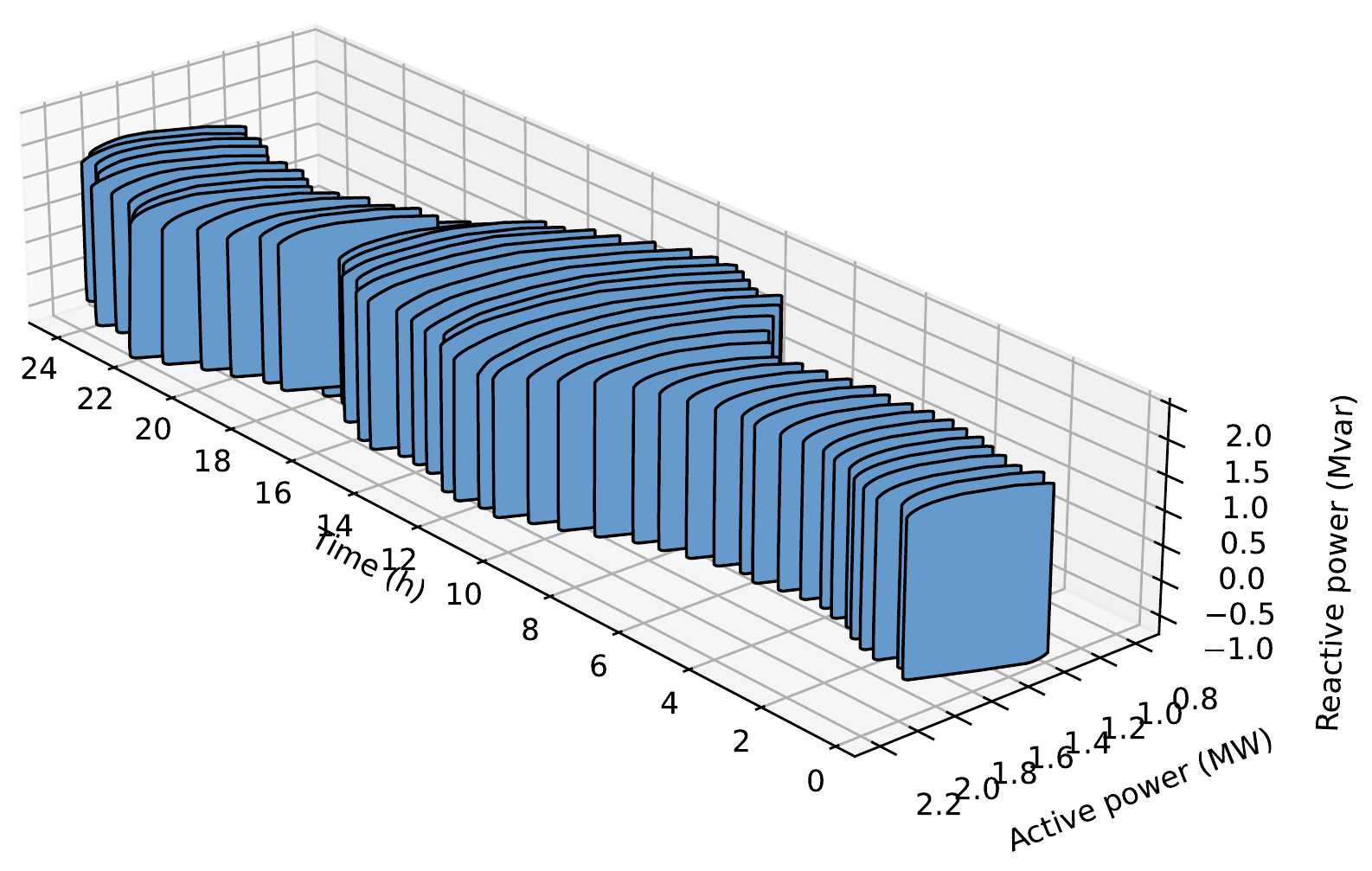}
	\caption{Time-series plots of the feeder-level flexibility region for a 24-hour period. (DERs considered include controllable loads, PV, and battery systems)}
    \label{fig_flex_timeseries_CL}
\end{figure}

\begin{figure}[!h]
	\centering
	\includegraphics[width=2.8 in]{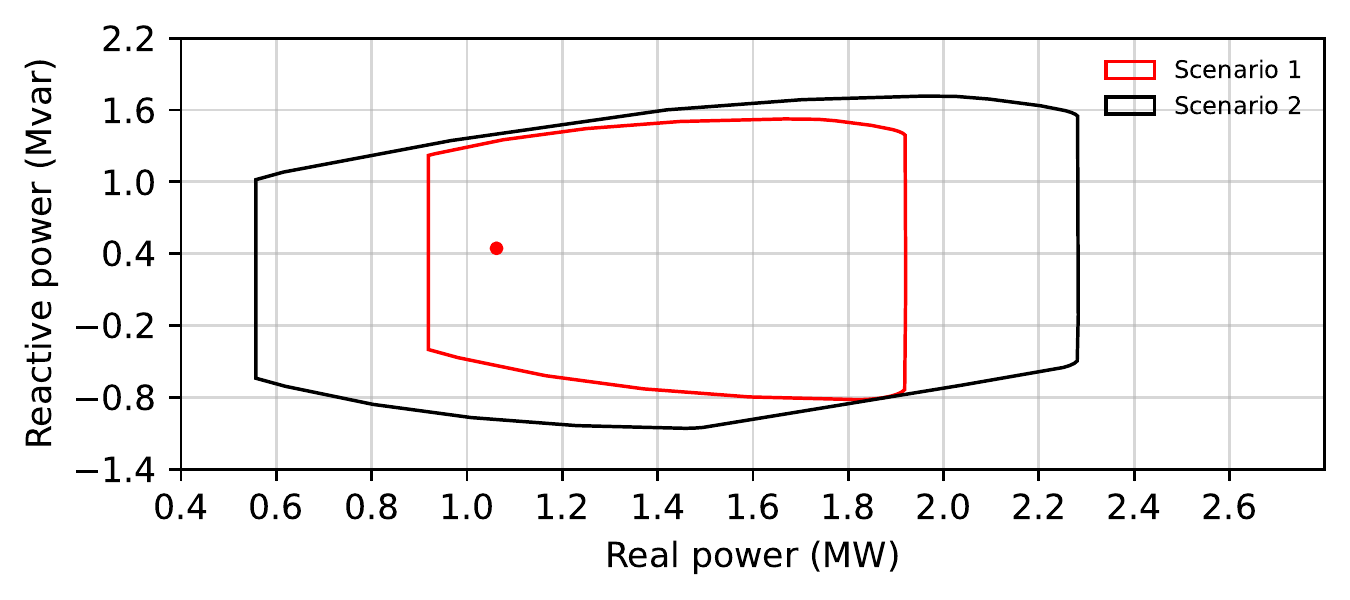}
	\caption{A side-by-side comparison of the feeder-level flexibility regions at 12:00 p.m. in a sunny day. (The red dot represents the feeder operating point).}
    \label{fig_t12_flexs}
\end{figure}

In the second scenario, controllable loads are included. All the load nodes in the 123-bus system (95 in total) are controllable. Note that the load feasible operating regions are defined in (\ref{equ_feasible_set_loads}). In this example, we set the lower and upper load active power limits as 0.8 and 1.2 times the scheduled power, respectively. The load reactive power consumption is calculated using the power factor originally defined in the IEEE 123-bus test system. As shown in Fig. \ref{fig_flex_timeseries_CL} and \ref{fig_t12_flexs}, controllable loads can significantly increase the feeder-level active power regulation range. However, if a controllable load cannot vary its power factor, its contribution to the reactive power regulation is very limited.

\subsection{Modeling PV and Load Forecast Errors}
In practice, flexibility region estimations are used to provide active and reactive power limits for energy scheduling algorithms (e.g., unit commitment and economic dispatch). The inputs are short-term load and PV forecast for subsequent intervals, the time steps of which can be 5, 15, or 30 minutes. Thus, accounting for load and PV forecast errors in the flexibility region estimation process is essential for the operators to quantify the uncertainty of using DERs to regulate the feeder active and reactive power injections. 

\begin{figure}[!h]
	\centering
	\includegraphics[width=2.3 in]{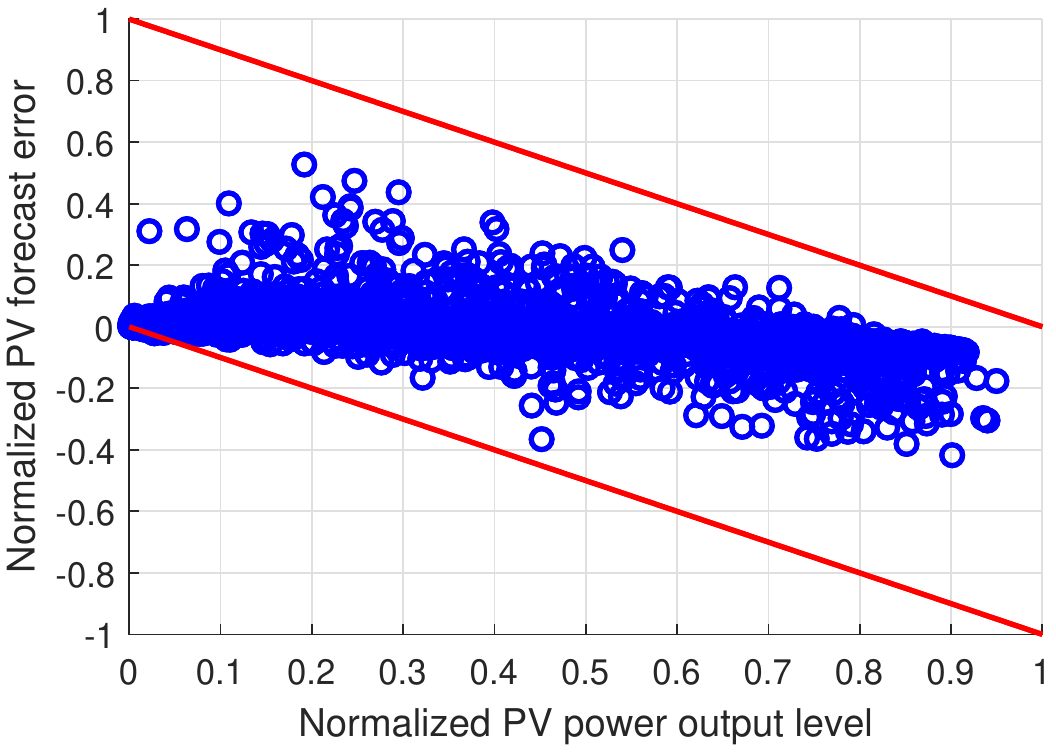}
	\caption{Normalized PV forecast errors at different PV power output levels.}
    \label{fig_PV_forecast_error_powerlevels}
\end{figure}

\begin{figure}[!h]
\centering
\subfloat[sunny]{\includegraphics[width=1.1 in]{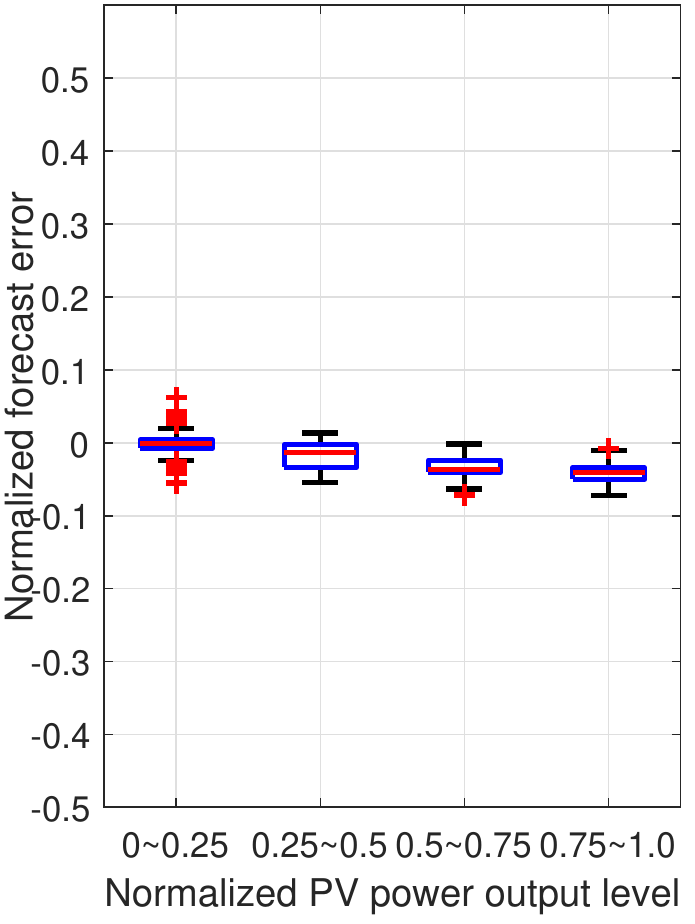}}%
\hfill
\subfloat[cloudy]{\includegraphics[width=1.08 in]{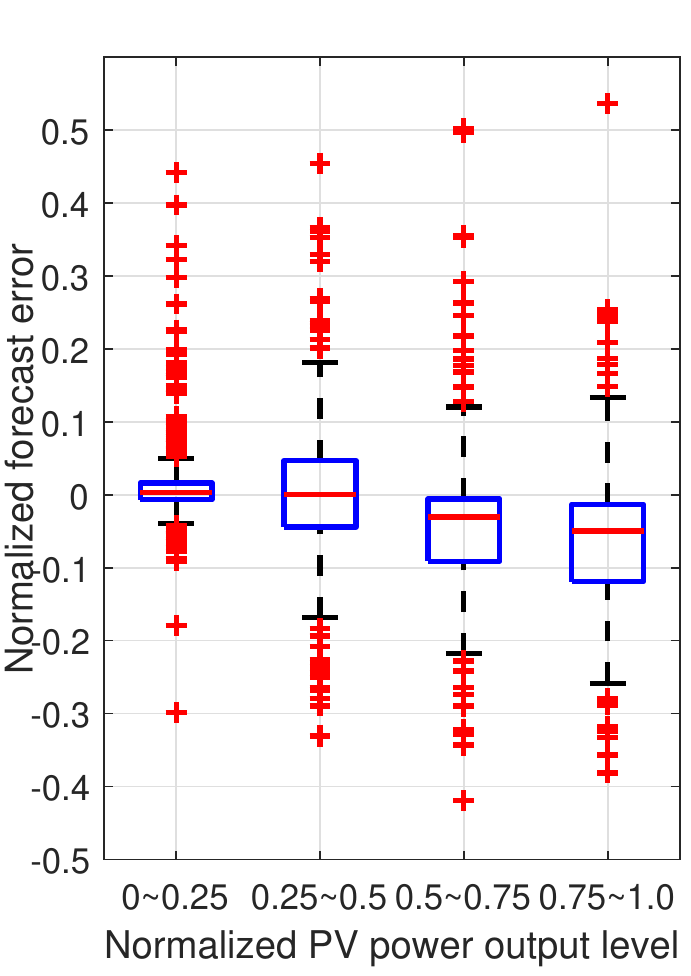}}
\hfill
\subfloat[overcast]{\includegraphics[width= 1.0 in]{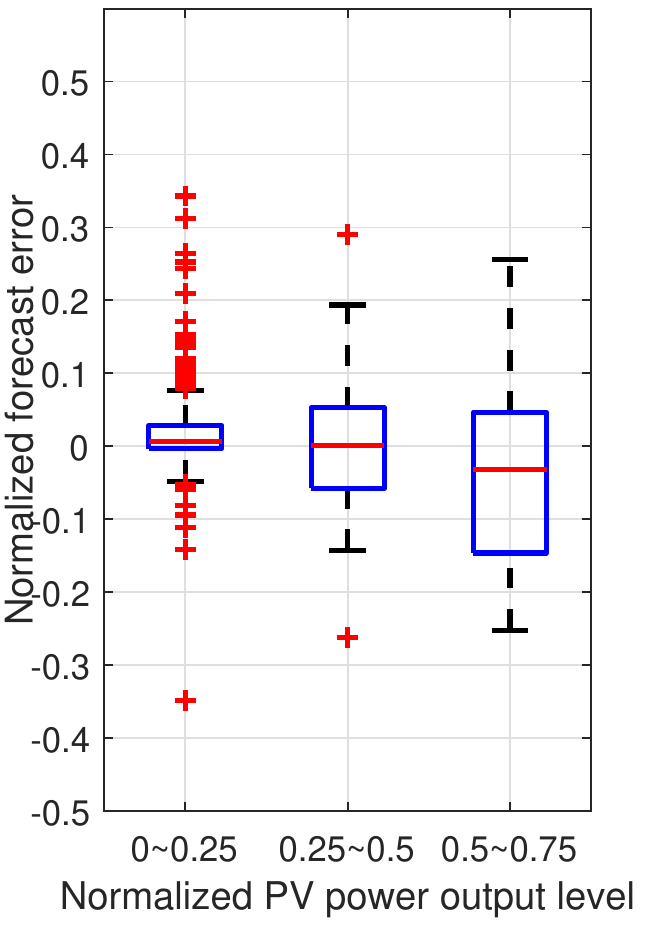}}
\caption{Box plots of the PV forecast errors in three weather conditions at four PV power output levels. (Note that in (c), which is the overcast case, the PV power outputs never reached 0.75-1.0 p.u.)}
\label{fig_PV_forecast_powerlevles}
\end{figure}

There are two sources of forecast errors: passive loads and PV outputs. In this paper, we assume that load forecast error follows Gaussian distribution. In\cite{vahedipour2020risk, baker2016energy}, the forecast error of a renewable generation resource is assumed to follow Gaussian distribution. However, in practice, the distributions of PV and wind forecast errors may be skewed and follow multimodal distributions \cite{zhao2019two}, as shown in the scattered plot of Fig. \ref{fig_PV_forecast_error_powerlevels}. Moreover, PV forecast accuracy is weather dependent, i.e., the forecast error is small on a sunny and is large on a cloudy day or overcast day, as shown in Fig. \ref{fig_PV_forecast_powerlevles}. 

Therefore, the scenario-based Gaussian mixture model (GMM) \cite{gu2018multi,bilil2018mmse} is used to find a mixture of multi-Gaussian probability distributions that best match the error data set. GMM likelihood optimization is used to fit GMMs using the iterative Expectation-Maximization (EM) algorithm\cite{singh2009statistical}. Analytic criteria, the Akaike information criterion (AIC) and the Bayesian information criterion (BIC)\cite{murphy2012machine}, are used as evaluation criteria to avoid underfitting and overfitting. The smaller the AIC and BIC values are, the better the fit will be.

\begin{figure}[!h]
	\centering
	\includegraphics[width=2.2 in]{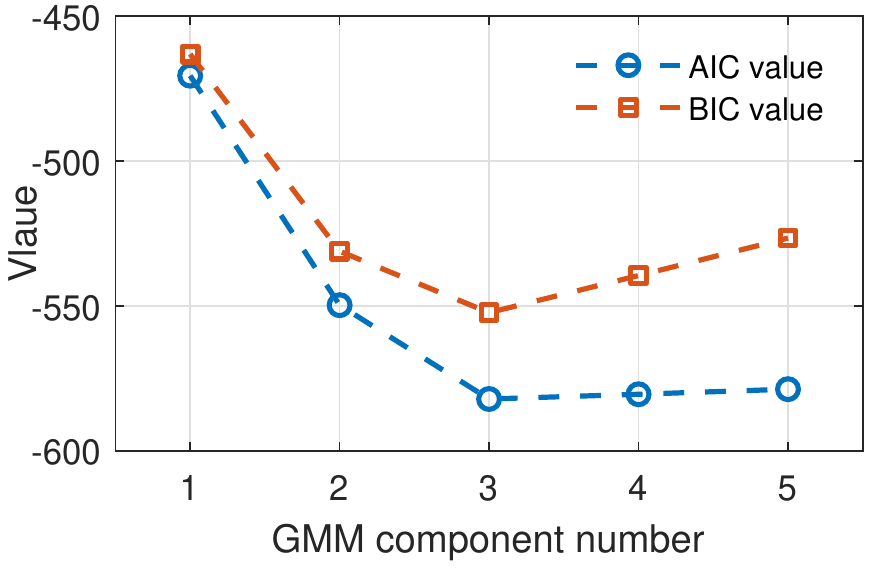}
	\caption{AIC/BIC value variations with different GMM component numbers.}
    \label{fig_GMM_components}
\end{figure}

\begin{figure}[!h]
	\centering
	\subfloat[PDF]{\includegraphics[width=.25\textwidth]{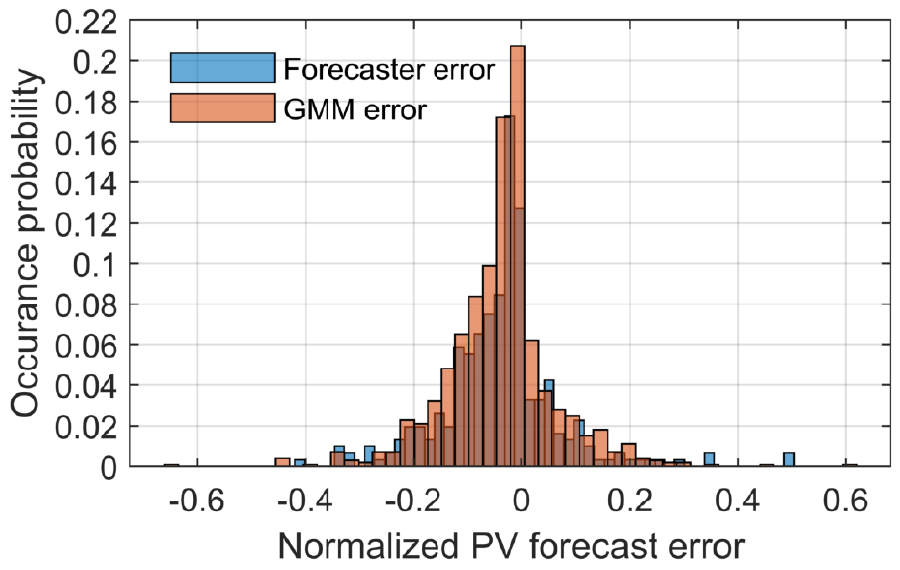}}
    \hfill
    \subfloat[CDF]{\includegraphics[width=.225\textwidth]{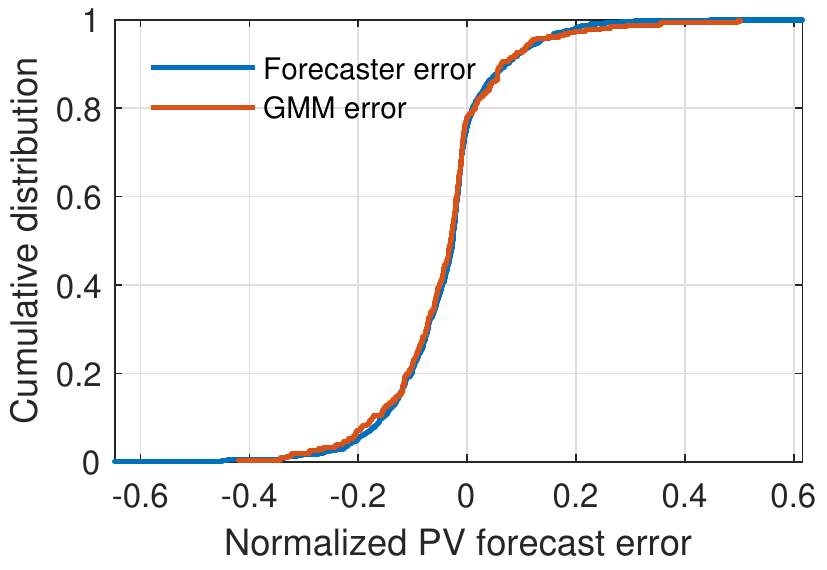}}
	\caption{A comparisons of error distribution between the actual and the fitted GMM generated errors. (Cloudy day, PV power level: 0.50-0.75 p.u)}
    \label{fig_GMM_comparision}
\end{figure}

To match the PDF of the PV forecast errors on a cloudy day at the power level of 0.50-0.75 p.u.,  the variations of AIC and BIC values for different component numbers are computed. As shown in Fig. \ref{fig_GMM_components}, the AIC and BIC values are the smallest when the component number is 3. Thus, three Gaussian distributions are needed to fit the PDF of the PV forecast errors. The fitted component weights are [0.4024, 0.1080, 0.4896]. The component Gaussian distributions are $\mathcal{N}\left(0.0024, 6.4572e^{-5}\right)$, $\mathcal{N}\left(0.0688, 0.0172\right)$, and $\mathcal{N}\left(0.0168, 9.4331e^{-4}\right)$.
As shown in Fig. \ref{fig_GMM_comparision}, the PDF and CDF of the GMMs-generated samples  match closely with those of the actual forecast errors. The same procedures are applied to determine the optimal component number for matching the PDFs of the PV forecast errors at different power levels for different day types.

\subsection{Uncertainty Propagation}
\begin{figure}[!h]
	\centering
	\includegraphics[width=2.65 in]{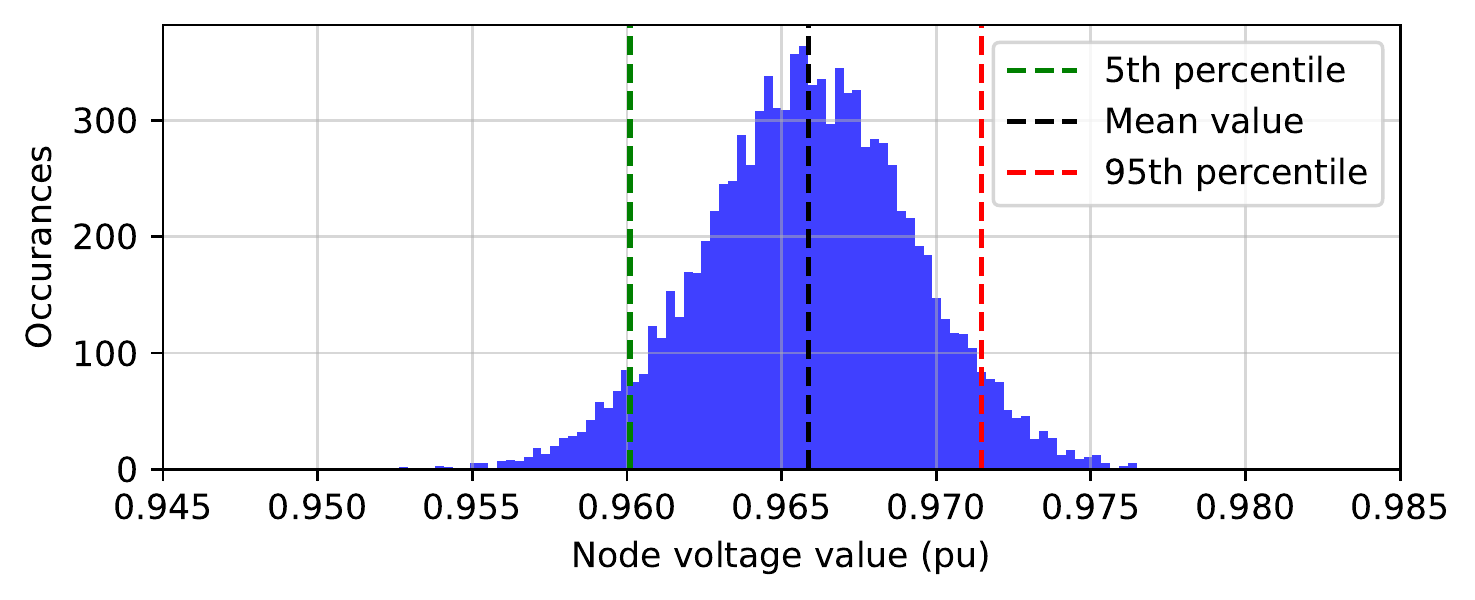}
	\caption{Distribution of voltage at a load node considering PV and load forecast errors.}
    \label{fig_node_voltage_distribution}
\end{figure}

\begin{figure}[!h]
	\centering
	\includegraphics[width=2.5 in]{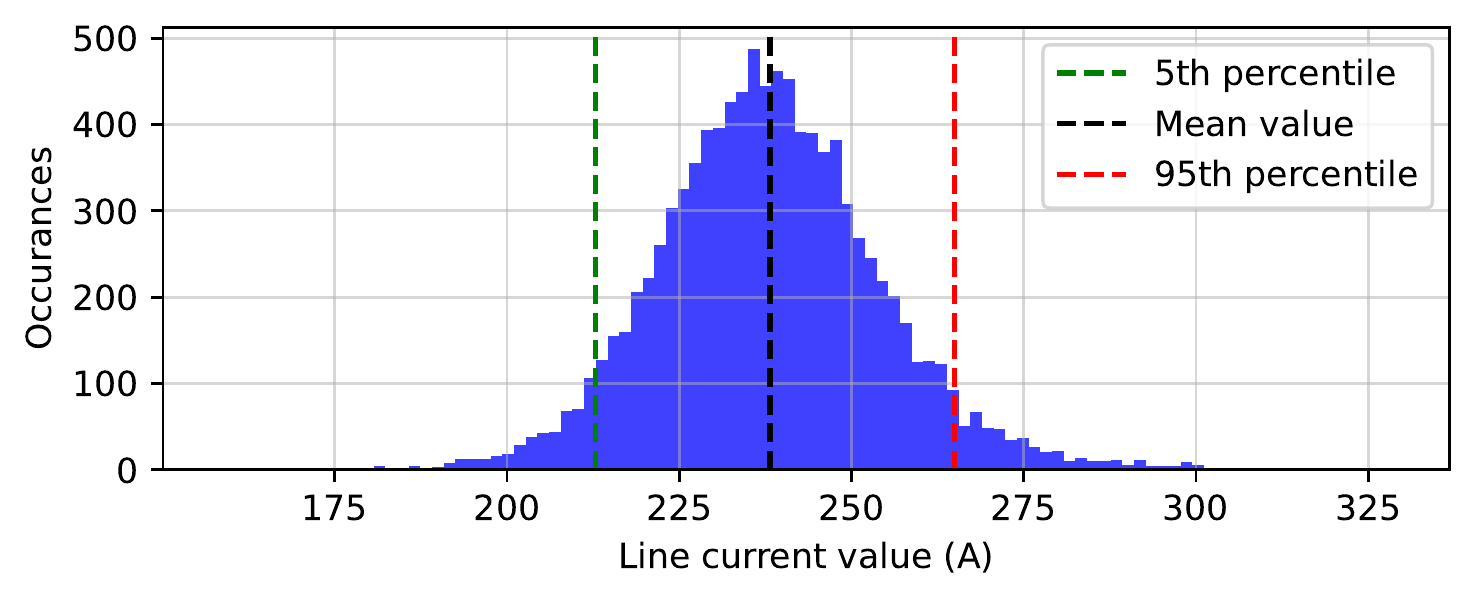}
	\caption{Distribution of current on a line considering PV and load forecast errors.}
    \label{fig_line_current_distribution}
\end{figure}

Uncertainties in PV and load forecasting introduce uncertainty in nodal power injection estimation, consequently making the power flow calculation an uncertainty propagation problem. Considering stochastic nodal injections in nonlinear AC power ﬂow calculation is time-consuming and hard to solve because it requires propagating uncertainties through a set of implicit nonlinear equations\cite{muhlpfordt2019chance}. To address this issue, the linearized power flow model introduced in section \ref{linearized_PF} (\ref{equ_linear_model_voltage}) and (\ref{equ_linear_model_current}) is used so the calculation of nodal voltage and current is linearized. Figs. \ref{fig_node_voltage_distribution} and \ref{fig_line_current_distribution} show an example of the resultant nodal voltage and line current value distribution respectively. With the uncertainties of nodal voltages and line flow currents modeled, the system-level operational chance constraints defined in (\ref{equ_system_cc_constraints}) can make sure system operational constraints are met with high probability in the flexibility  aggregation process.

\subsection{Impacts of Uncertainty on Flexibility Region Estimation}
Six simulation cases are presented to demonstrate the proposed DRCC method and the impacts of forecasting errors. The flexibility region at 12:00 p.m. is selected to illustrate the impacts. This is because the feeder load valley typically occurs at noon on a sunny day based on our case setup and the trends of the flexibility region variation is similar for other time slots as shown in Figs. \ref{fig_flex_timeseries} and \ref{fig_flex_timeseries_CL}. In addition, to explicitly show the impacts of the uncertainty in PV forecasting errors, the DERs considered do not include controllable loads, i.e., only the flexibility of 9 PVs and 1 BESS are considered. Note that in Figs. \ref{fig_flex_timeseries_CL} and \ref{fig_t12_flexs}, we have shown that the controllable loads mainly affect the real power regulation capability and its reactive power regulation capabilities is negligible.



\subsubsection{Device-level Chance Constraints}
\begin{figure*}[h]
\centering
\subfloat[\label{fig_flex_PV_confidence1}]{\includegraphics[width= .31\textwidth]{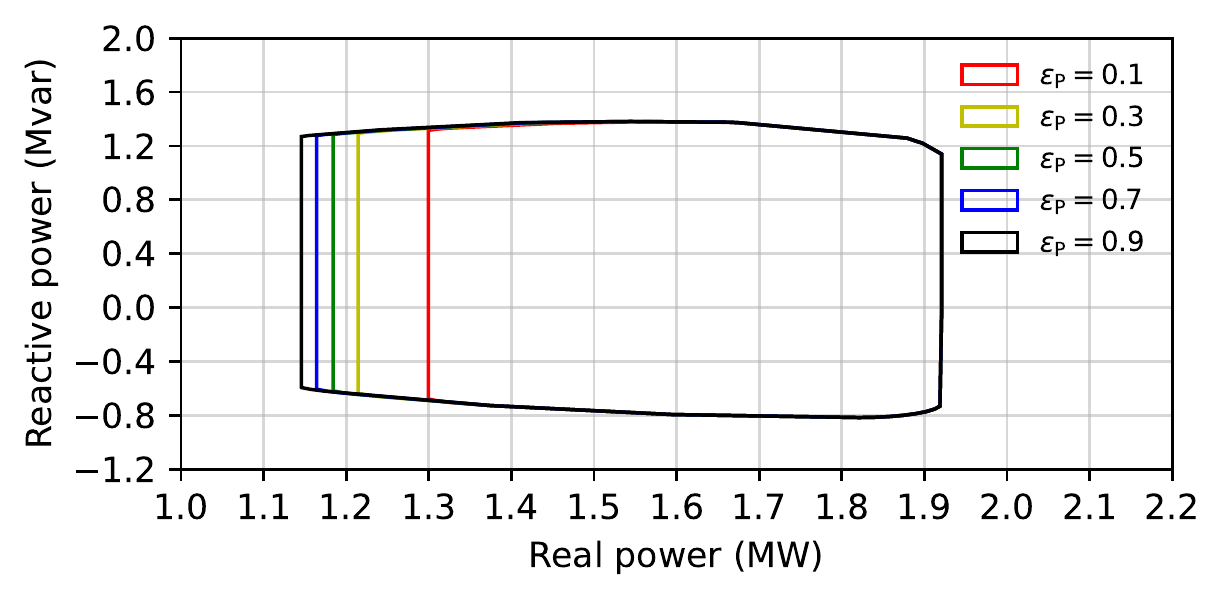}}
\hfill
\subfloat[\label{fig_flex_PV_confidence2}]{\includegraphics[width= .35\textwidth]{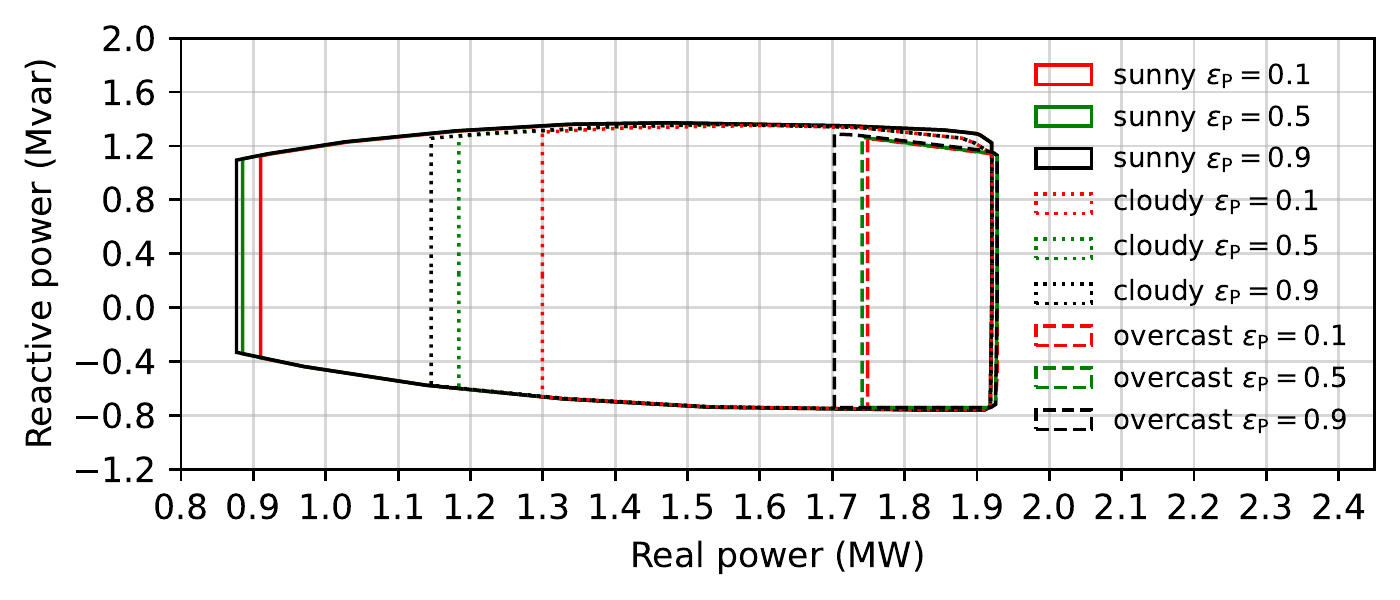}}
\hfill
\subfloat[\label{fig_flex_IEEE1547}]{\includegraphics[width= .31\textwidth]{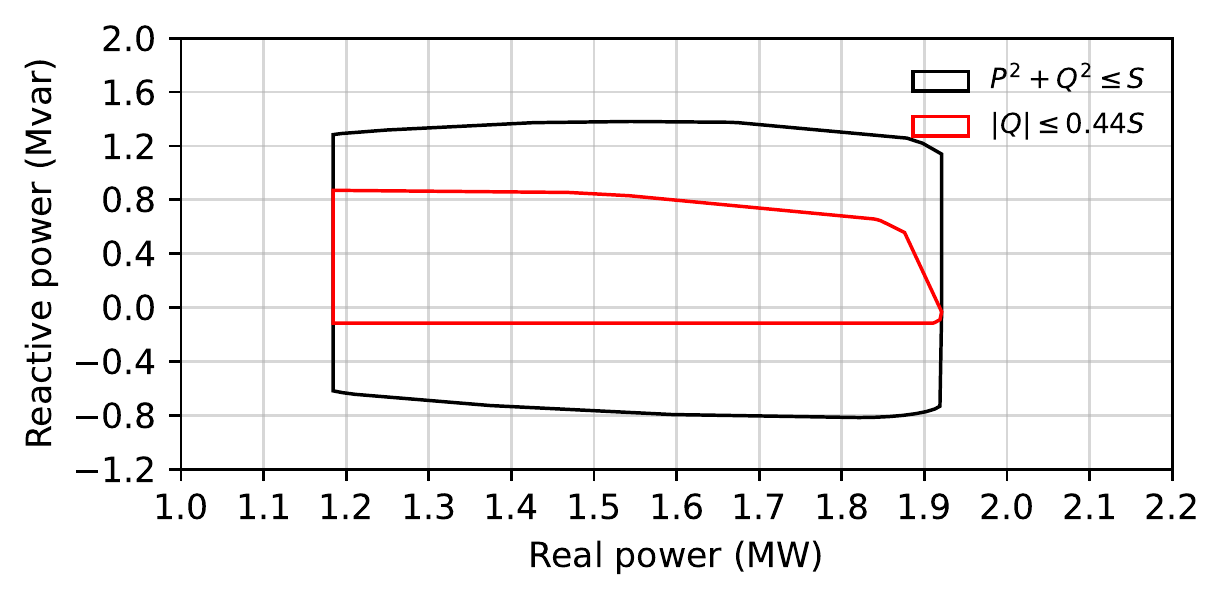}}
\hfill\\
\subfloat[\label{fig_flex_I_confidence}]{\includegraphics[width= .31\textwidth]{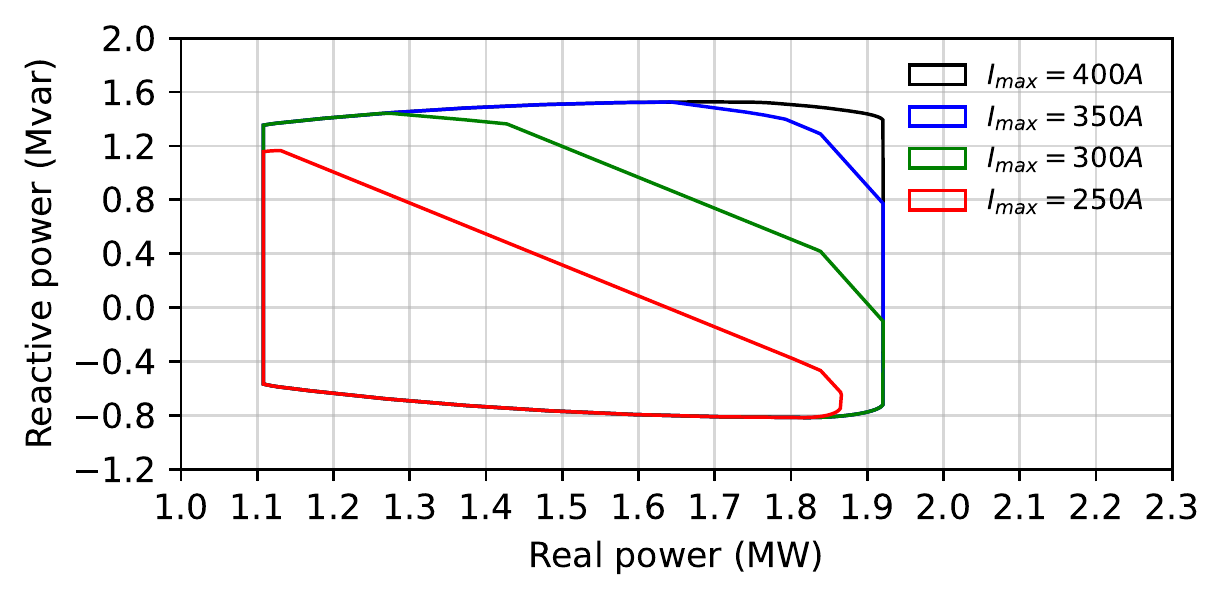}}
\hfill
\subfloat[\label{fig_flex_V_confidence}]{\includegraphics[width= .35\textwidth]{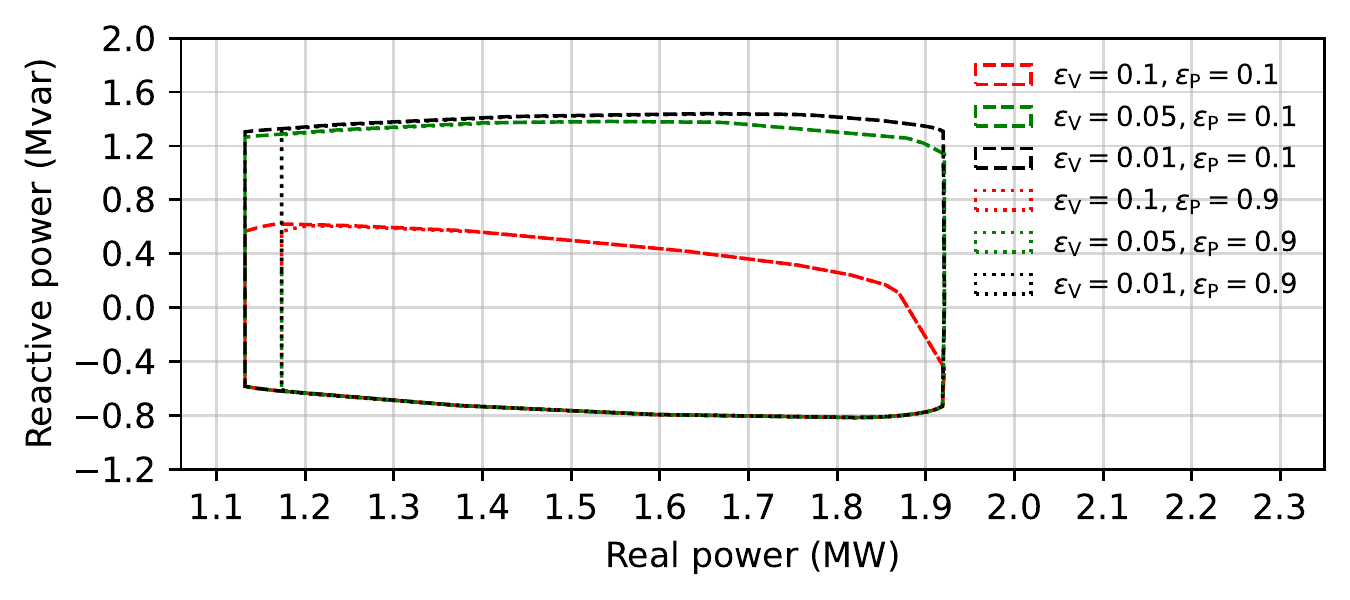}}
\hfill
\subfloat[\label{fig_flex_search_directions}]{\includegraphics[width= .31\textwidth]{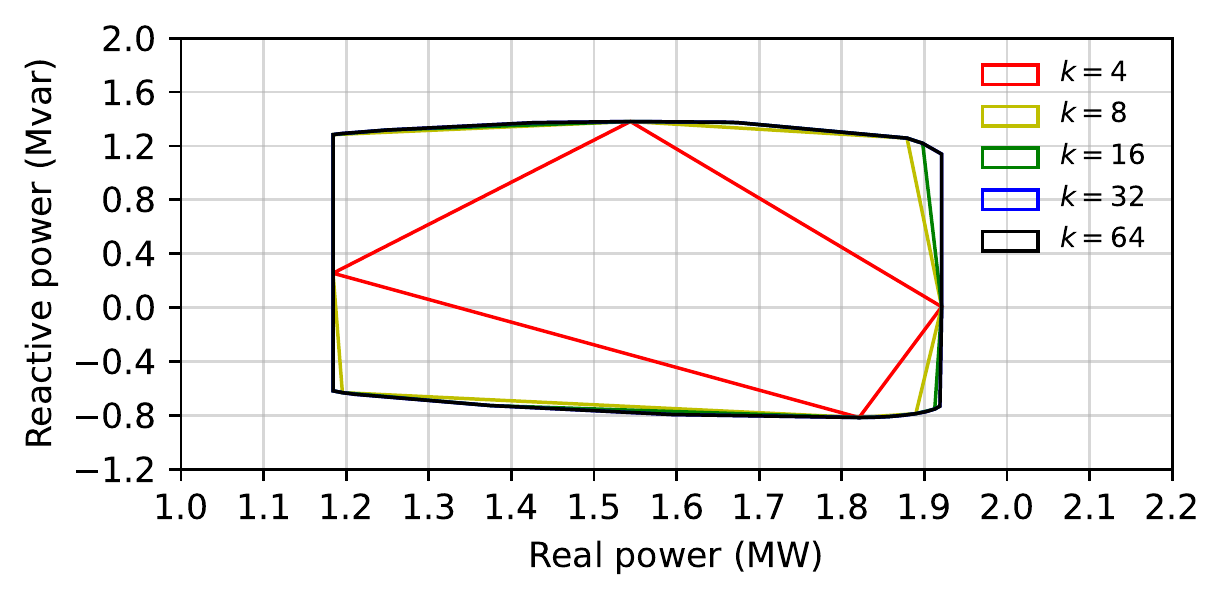}}
\caption{An illustration of the main factors that affect the DER flexibility region aggregation: (a) device-level risk factor, $\epsilon_{\mathrm{P}}$, (b) different day types, (c) inverter reactive power regulation limits, (d) line flow current limits, (e) nodal voltage limit violation probabilities, $\epsilon_{\mathrm{V}}$, and (f) the number of search direction.  Note that: i) for cases (a), (c)-(f), the day type is cloudy, ii) all flexibility regions are calculated at 12:00 p.m., and iii) in the figures, "real power" refers to the feeder-head real power consumption, $\tilde{P}_{t}^{\mathrm{F}}$, and "reactive power" refers to the feeder-head reactive power consumption, $\tilde{Q}_{t}^{\mathrm{F}}$. } 
\label{fig_DRCC}
\end{figure*}

At the device level, feasible  region  violation  probability $\epsilon_{\mathrm{P}}$ is used to quantify the confidence level of the estimated flexibility region boundary. 
On a cloudy day, as shown in Fig. \ref{fig_DRCC} (a), the active power regulation capability decreases when $\epsilon_{\mathrm{P}}$ decreases because more conservative estimations of the PV real power outputs are used. Meanwhile, the reactive power regulation capability remains the same. This is because the PV reactive power regulation range is insensitive to the irradiance variations.
Note that in all five different $\epsilon_{\mathrm{P}}$ scenarios, the upper limit of the feeder-level real power consumption does not change with respect to $\epsilon_{\mathrm{P}}$.  This is because the feeder reaches its highest real power consumption level when all the PV systems connected to the feeder are turned off. Thus, its upper real power consumption limit is insensitive to PV forecast errors.

In Fig. \ref{fig_DRCC} (b), the flexibility regions of sunny, cloudy, and overcast days with different $\epsilon_{\mathrm{P}}$ settings are plotted. As expected, in sunny and overcast days, the flexibility region variations with $\epsilon_{\mathrm{P}}$ are very small while in a cloudy data the variation can be significant.  Thus, the selection of different $\epsilon_{\mathrm{P}}$ will have greater impacts on the active power regulation capability in a cloudy day. Selecting a small $\epsilon_{\mathrm{P}}$ may reduce the risk of non-compliance actions when providing grid services.  

There are other factors that can influence the DER real and reactive power regulation capabilities. For example, the new IEEE 1547-2018 DER integration standard\cite{IEEE1547_2018} requires that the minimum reactive power capability of the DER be set as 44\% of the inverter kVA rating. Compared with the two-dimensional flexibility region defined in (\ref{equ_feasible_set_PV}), if the maximum reactive power output level is set as 44\% of the inverter rated power, the device-level flexibility region will shrink significantly, especially in dimension of reactive power regulation capability, as shown in Fig. \ref{fig_DRCC} (c). 

\subsubsection{Line Current Chance Constraints}
The system-level operational chance constraints of the line current limits are defined in (\ref{equ_system_cc_constraints_Imax}). When $\epsilon_{\mathrm{P}}=0.5$  and $\epsilon_{\mathrm{V}}= 0.05$, the variations of aggregated flexibility region for four different maximum line flow current $I^{\mathrm{max}}_{l}$ settings in a cloudy day are shown in Fig. \ref{fig_DRCC} (d). When $I^{\mathrm{max}}_{l} = 400$ A, the size of the flexibility region is the same as that of the no-current-limit case. When $I^{\mathrm{max}}_{l}$ decreases from 400 A to 250 A, the flexibility region will shrink quickly in both the real and reactive power regulation directions. 

\subsubsection{Nodal Voltage Chance Constraints}
The flexibility region aggregation is also constrained by the nodal voltage operation limits defined in (\ref{equ_system_cc_constraints_Vmax}) and (\ref{equ_system_cc_constraints_Vmin}). The flexibility region accounting for the probabilities of voltage violation  $\epsilon_{\mathrm{V}}$ when $\epsilon_{\mathrm{P}}$ is 0.1 and 0.9 for a cloudy day, are shown in Fig. \ref{fig_DRCC} (e).  The results show that the size of the flexibility region when $\epsilon_{\mathrm{V}}=0.01$ is much smaller than that of the case when $\epsilon_{\mathrm{V}}=0.05$ or $\epsilon_{\mathrm{V}}=0.1$. This shows that the flexibility region becomes smaller to achieve higher confidence levels of meeting the nodal voltage operation limits. Moreover, the active power regulation capability decreases when $\epsilon_{\mathrm{P}}$ decreases, because the estimations of PV real power outputs become more conservative correspondingly.

\subsubsection{Number of Search Directions}
\begin{figure}[!h]
	\centering
	\includegraphics[width=2.5 in]{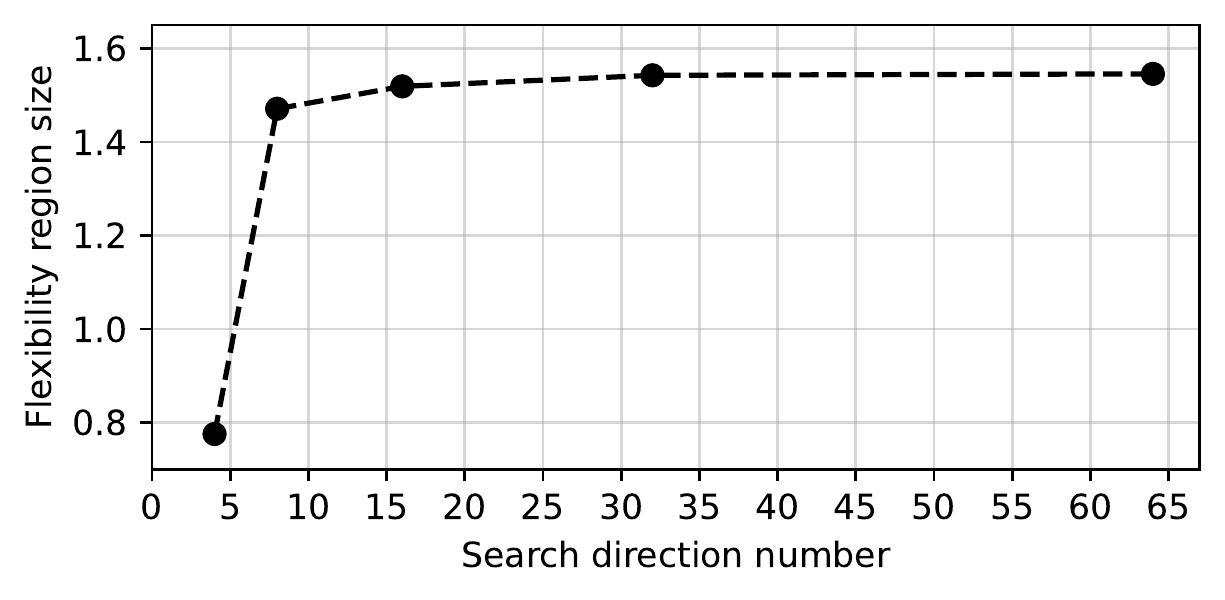}
	\caption{Aggregated flexibility region size with different search directions.}
    \label{fig_flex_size_variations}
\end{figure}

The aggregated flexibility region can be represented by a polygon as shown in Fig. \ref{fig_search_direction}. When we increase the number of search directions, $k$, the approximated flexibility region will become closer to the actual one, as shown in Fig. \ref{fig_DRCC} (f). The area of the polygon-based, two-dimensional (P\&Q) flexibility region $S$ can be calculated using the  shoelace formula\cite{Braden1986}. As shown in Fig. \ref{fig_flex_size_variations}, when the search direction number $k$ increases, the area of the flexibility region size will increase. However, when $k>16$, the increase of the area covered by the flexibility region is negligible, showing that the 16 or 32-edge polygon is sufficient for approximating the flexibility region.



\subsection{Computing Time of the DRCC Algorithm }\label{sensitivity_analysis}
The DRCC algorithm (\ref{equ_DRCC}) needs to be solved $k$ times to identify the bounding points on the $k$ search directions, which define the flexibility region boundaries. The second-order cone programming-based reformulation allows the optimization problem to be solved quickly using off-the-shelf solvers. In our case, an open-source package, PICOS, is used.  PICOS provides a Python interface to the conic optimization solvers \cite{sagnol2012picos}. The DRCC algorithm is solved using Gurobi solver\cite{gurobi} on a desktop with Intel Core i7 CPU @ 3.40 GHz and 16 GB of RAM. For $k=32$, the computation time for the flexibility region estimation is between 15.2 seconds to 19.7 seconds for multiple runs. This shows that the DRCC algorithm can meet the real-time operation runtime requirement.

\section{Conclusion}\label{section_conclusion}
Identifying the flexibility region for aggregated distributed controllable resources and quantifying the risk in their operation are essential for using DERs to provide grid services.  Visualizing the progressive change in the flexibility region during the scheduling period is also crucial for the safe and reliable operation of the distribution grid with high-penetration of DERs. The proposed data-driven, distributionally robust chance-constrained optimization method can aggregate the device-level DER flexibility region to the feeder-level considering the forecasting errors and the physical coupling between the real and reactive power limits of each DER. By running  fixed-point  linearized  power flow studies, the distribution nodal voltage and line current operational limits can also be accounted for to ensure the feasibility of the control actions that are taken within the flexibility region. The risks of controlling DERs for providing grid services are quantified by considering the forecasting errors of the nodal power injections and the propagation of such errors in the flexibility aggregation process. Simulation results demonstrate the impacts of uncertainty on DER operation limits and the effectiveness of the probabilistic-based flexibility region derivation.  The runtime statistics show that the computing speed of the proposed DRCC algorithm meets real-time operation requirements.

\appendices
\section*{Acknowledgment}
The authors would like to thank Dr. David Lubkeman at NC State University, James Stoupis, David Coats, and 	
Mohammad Rrazeghi-jahromi with ABB for their support and comments. 

\ifCLASSOPTIONcaptionsoff
  \newpage
\fi




\bibliographystyle{IEEEtran}
\bibliography{Flexibility_Aggregation.bib}

\end{document}